\documentclass[11pt]{article}
\usepackage[latin1]{inputenc} 
\usepackage{listings, tcolorbox}

\usepackage{}
\usepackage{amssymb}
\usepackage{amsfonts}
\usepackage{mathrsfs,psfrag,eepic,epsfig}
\usepackage{graphicx}
\usepackage{amsmath}
\usepackage{color}
\usepackage{cite}
\usepackage{amsthm}
\usepackage{txfonts}
\usepackage{pifont,bm}
\usepackage{tikz}
\usepackage{epstopdf}
\usepackage{enumerate}
\usepackage{lineno,hyperref}
\modulolinenumbers[5]

\newtheorem{thm}{Theorem}[section]
\newtheorem{cor}[thm]{Corollary}

\newtheorem{prop}[thm]{Proposition}
\theoremstyle{definition}

\newtheorem{rem}[thm]{Remark}

\makeatletter
\def\@biblabel#1{[#1]}
\makeatother

\makeatletter \@addtoreset{equation}{section}
\renewcommand{\theequation}{\arabic{section}.\arabic{equation}}

\begin{document}

\begin{titlepage}
\title{\bf{Riemann-Hilbert problem for the nonlinear Schr\"{o}dinger equation with multiple high-order poles under nonzero boundary conditions
\footnote{
Corresponding author.\protect\\
\hspace*{3ex} \emph{E-mail addresses}:  sftian@cumt.edu.cn,
shoufu2006@126.com (S. F. Tian)}
}}
\author{Jin-Jie Yang, Shou-Fu Tian$^{*}$ and Zhi-Qiang Li\\
\small \emph{School of Mathematics, China University of Mining and Technology, Xuzhou 221116,}\\
\small \emph{ People's Republic of China}
\date{}}

\thispagestyle{empty}
\end{titlepage}
\maketitle

\vspace{-0.5cm}
\begin{center}
\rule{15cm}{1pt}\vspace{0.3cm}

\parbox{15cm}{\small
{\bf Abstract}\\
\hspace{0.5cm}  The Riemann-Hilbert (RH) problem is first developed to study the focusing nonlinear Schr\"{o}dinger (NLS) equation with multiple high-order poles under nonzero boundary conditions. Laurent expansion and Taylor series are employed to replace the residues at the simple- and the second-poles. Further, the solution of RH problem is transformed into a closed system of algebraic equations, and the soliton solutions corresponding to the transmission coefficient $1/s_{11}(z)$  with an $N$-order pole  are obtained by solving the algebraic system. Then, in a more general case, the transmission coefficient with multiple high-order poles is studied, and the corresponding solutions are obtained. In addition, for high-order pole, the propagation behavior of the soliton solution corresponding to a third-order pole is given as example.}

\vspace{0.5cm}
\parbox{15cm}{\small{

\vspace{0.3cm} \emph{Key words:} The focusing nonlinear Schr\"{o}dinger equation; Riemann-Hilbert problem; Nonzero boundary conditions; Multiple high-order poles.\\
}
}
\end{center}
\vspace{0.3cm} \rule{15cm}{1pt} \vspace{0.2cm}



\section{Introduction}
The solution of nonlinear partial differential equation plays an important role in nonlinear science, because nonlinear wave can describe many physical phenomena, for example nonlinear Schr\"{o}dinger (NLS) equation can describe the evolution of complex envelope of weakly nonlinear dispersive wave trains. After years of development, many effective methods have been proposed to solve nonlinear differential equations. One of them is called inverse scattering transformation (IST), which was proposed by Gardner, Greene, Kruskal and Miura for studying Korteweg-de Vries (KdV) equation with fast decay initial value \cite{GGKM}. In 1972, Zakharov and Shabat extended this idea to the initial value problem of NLS equation \cite{ZS-1972}, and then Manakov, Ablowitz and Newell \emph{et al.} further extended this method to study more abundant nonlinear equations with initial value problems \cite{Ablowitz-PRL1973,Ablowitz-SAPM1974,Manakov-1974,Ablowitz-JMP1973,Ablowitz-1981}.

A modern version of IST, the so-called Riemann-Hilbert (RH) problem \cite{BC-1984,Zhou-1989}, was proposed.  Compared with IST, an important difference is that the reconstruction formula of the solution to nonlinear differential equation is recovered by the solution of the corresponding RH problem, rather than the Gel'fand-Levitan-Marchenko integral equation, which greatly simplifies the complexity of the analysis process.  In addition, RH problem can also be used to solve the problems related to higher-order spectrum, thus RH problem has gradually become a powerful tool to solve the initial value and initial boundary value problems of integrable systems, such as the coupled NLS equation \cite{Tian-jde,Tian-pa2016,YZhang-2020}, the coupled modified KdV equation \cite{Tian-jpa2017,Ma-2018,NLiu-2019}, derivative nonlinear Schr\"{o}dinger equation \cite{tian-PA2018}, and Sasa-Satsuma equation \cite{Geng-2016,Xu-2013} etc. It is noted that by using RH method to construct exact solutions of nonlinear differential equations, many literatures require that the initial values belong to Schwarz space, that is, the initial values tend to zero at infinity.

Recently, Biondini and Kovai improved the work in \cite{Ma-1979} and proposed a  rigorous framework for studying soliton solutions of the focusing NLS equation in \cite{Biondini-JMP2014}
\begin{align}\label{Q1}
iq_{t}+q_{xx}+2(|q|^{2}-q_{0}^{2})=0,
\end{align}
under the nonzero boundary conditions (NZBCs)
\begin{align}\label{Q2}
\lim_{x\rightarrow\pm\infty}q(x,t)=q_{\pm},
\end{align}
where $|q_{\pm}|=q_{0}\neq0$. Many scholars have extended this method to other integrable models, including  mKdV (local/nonlocal) equation \cite{Zhang-2020,Zhang-2020-non}, modified NLS equation \cite{Yang-2021}, Sasa-Satsuma equation \cite{Wll}, modified Landau-Lifshitz equation \cite{Yang-TMP} and Gerdjikov-Ivanov (GI) equation \cite{GI-2020}.  It is known that for nonlinear integrable systems, soliton solutions are generated at the poles of transmission coefficients, and most of integrable systems have higher-order poles. It should be noted that the solution of GI equation with multiple higher order poles is obtained by introducing Laurent series and Taylor expansion in \cite{GI-NZBC}. However, Biondini and Kovai mainly discussed the soliton solutions to the NLS equation \eqref{Q1} of the transmission coefficient with simple poles under NZBCs in \cite{Biondini-JMP2014}. In 2017, Pichler and Biondini further studied the transmission coefficient with second-order poles in \cite{Pichler-2017}. In this case, the residue corresponding to the negative second power of the discrete spectrum needs to be considered, which leads to complicated calculation.  Theoretically, for the $N$-order poles, we only need to calculate the corresponding residues, then the potential function can be restored by  the regularized RH problem via subtracting the residues at the poles and the asymptotic behavior (Note the regularized RH can be solved by the Plemelj formula).

For multiple higher-order poles in integrable systems, the method of merging multiple poles is generally adopted. The solution of modified KdV equation with second and third order poles was investigated in \cite{Wadati-1982}. By coalescing different simple poles, Tsuru and Wadati studied the $N$-order pole solutions of the sine-Gordon equation \cite{Tsuru-1984}. It should be noted that the combination of poles can not guarantee that the solution is regular, and it also involves a lot of limit calculation for multiple high-order poles solutions, so it is not feasible. The dressing method and Darboux transformation are often used to deal with higher-order pole soliton solutions, including NLS equation \cite{Gagnon-1994} and $N$-wave system \cite{Shchesnovich-JMP,Shchesnovich-SAMP}. Recently, the high-order pole solution and the regularity of the multiple high-order poles solitons under zero boundary conditions are investigated by using Laurent expansion and Taylor expansion to avoid the calculation of each order poles in \cite{Zhang-SAMP,Zhang-2019}. Therefore, this work is to generalize the results presented in \cite{Biondini-JMP2014,Pichler-2017} under NZBCs, and obtain an $N$-order pole soliton solution and multiple high-order poles soliton solutions of the focusing NLS equation \eqref{Q1}.

The frame of the work is arranged as: In section 2, the conditions needed to establish the corresponding RH problem and the establishment of RH problem are briefly described. The case of transmission coefficient with a higher-order pole is discussed in section 3, and the corresponding soliton solution expression is obtained, which can be degenerated to the results in \cite{Biondini-JMP2014,Pichler-2017} with special parameters. The case of transmission coefficient with multiple higher-order poles is discussed, and the concrete expression of the solution is given in section 4. Finally some conclusions and discussions are presented in the last section.

\section{Riemann-Hilbert problem for the focusing NLS equation}
\subsection{Spectrum analysis}
The NLS equation \eqref{Q1} is of the Lax pair
\begin{align}\label{Q3}
\phi_{x} =X\phi ,\qquad\phi_{t} =T\phi ,
\end{align}
where $\sigma_{3}=\left(
                    \begin{array}{cc}
                      1 & 0 \\
                      0 & -1 \\
                    \end{array}
                  \right)$, $Q=\left(
                                 \begin{array}{cc}
                                   0 & q \\
                                   -q^{*} & 0 \\
                                 \end{array}
                               \right)$,~~$X=(ik\sigma_{3}+Q)$,~~
                               $T=-2kX+i\sigma_{3}\left(Q_{x}-Q^{2}-q_{0}^{2}\right)$.

Under the NZBCs \eqref{Q2}, the asymptotic spectra problem of the Lax pair \eqref{Q3} reads
\begin{align}\label{Q4}
\phi_{x}=X_{\pm}\phi,\quad \phi_{t}=T_{\pm}\phi,
\end{align}
where $X_{\pm}=\lim_{x\rightarrow\infty}X=ik\sigma_{3}+Q_{\pm}$, and $T_{\pm}=\lim_{x\rightarrow\infty}T=
-2kX_{\pm}$. It is easy to check that $\pm i\sqrt{k^{2}+q_{0}^{2}}$ are the eigenvalues of the  matrix $X_{\pm}$, note that these eigenvalues are multi-value functions, and the two branch points are $\pm iq_{0}$. Then the two-sheeted Riemann surface shown in figure 1 defined by $\lambda=\sqrt{k^{2}+q_{0}^{2}}$ is introduced to make sure the function $\lambda(k)$ is a single-value function on each sheet. Note that $\lambda(k)=\text{sign}(k)\sqrt{k^{2}+q_{0}^{2}}$ on the sheet $S_{1}$ and $\lambda(k)=-\text{sign}(k)\sqrt{k^{2}+q_{0}^{2}}$ on the sheet $S_{2}$. Since the introduction of Riemann surface transforms the complex $k$-plane into two-sheeted Riemann surface, the problem is still complicated. \\

\centerline{\begin{tikzpicture}[scale=0.5]
\draw[-][thick](1,5)--(13,5);
\draw[-][thick](13,5)--(13,-5);
\draw[-][thick](13,-5)--(1,-5);
\draw[-][thick](1,-5)--(1,5);
\draw[-][thick](-1,5)--(-13,5);
\draw[-][thick](-13,5)--(-13,-5);
\draw[-][thick](-13,-5)--(-1,-5);
\draw[-][thick](-1,-5)--(-1,5);
\draw[-][thick](-13,0)--(-7,0);
\draw[->][thick](-7,0)--(-1,0);
\draw[-][thick](1,0)--(7,0);
\draw[->][thick](7,0)--(13,0);
\draw[->][thick](7,0)--(7,5);
\draw[-][thick](7,0)--(7,-5);
\draw[->][thick](-7,0)--(-7,5);
\draw[-][thick](-7,0)--(-7,-5);
\draw[-][red, dashed](6,2)--(8,2);
\draw[-][red, dashed](6,-2)--(8,-2);
\draw[-][red, dashed](-6,2)--(-8,2);
\draw[-][red, dashed](-6,-2)--(-8,-2);
\draw[-][dashed](7,2)--(12,4);
\draw[-][dashed](7,0)--(12,4);
\draw[-][dashed](7,-2)--(12,4);
\draw[-][dashed](-7,2)--(-2,4);
\draw[-][dashed](-7,0)--(-2,4);
\draw[-][dashed](-7,-2)--(-2,4);
\draw[fill] (4,4) node{\tiny{$\lambda=-[(k-iq_{0})(k+iq_{0})]^{1/2}$}};
\draw[fill] (-10,4) node{\tiny{$\lambda=[(k-iq_{0})(k+iq_{0})]^{1/2}$}};
\draw[fill] (7,2) circle [radius=0.08];
\draw[fill] (7,-2) circle [radius=0.08];
\draw[fill] (12,4) circle [radius=0.06];
\draw[fill] (-2,4) circle [radius=0.06];
\draw[fill] (-7,-2) circle [radius=0.08];
\draw[fill] (-7,2) circle [radius=0.08];
\draw[fill] (7,2.2) node[left]{\tiny{$iq_{0}$}};
\draw[fill] (7,-1.8) node[left]{\tiny{$-iq_{0}$}};
\draw[fill] (7.3,-1.8) node[right]{\tiny{$\theta_{2}$}};
\draw[fill] (7.3,0.2) node[right]{\tiny{$\theta$}};
\draw[fill] (7.3,2.2) node[right]{\tiny{$\theta_{1}$}};
\draw[fill] (12,4) node[above]{\tiny{$k=re^{i\theta}$}};
\draw[fill] (-2,4) node[above]{\tiny{$k=re^{i\theta}$}};
\draw[fill] (-6.7,-1.8) node[right]{\tiny{$\theta_{2}$}};
\draw[fill] (-6.7,0.2) node[right]{\tiny{$\theta$}};
\draw[fill] (-6.7,2.2) node[right]{\tiny{$\theta_{1}$}};
\draw[fill] (-7,2.2) node[left]{\tiny{$iq_{0}$}};
\draw[fill] (-7,-1.8) node[left]{\tiny{$-iq_{0}$}};
\draw[fill] (12.5,0) node[below]{\tiny{$Re~k$}};
\draw[fill] (7,4.7) node[right]{\tiny{$Im~k$}};
\draw[fill] (-7,4.7) node[right]{\tiny{$Im~k$}};
\draw[fill] (-1.8,0) node[below]{\tiny{$Re~k$}};
\filldraw[red, line width=1.5] (7,2) to (7,-2);
\filldraw[red, line width=1.5] (-7,2) to (-7,-2);
\draw[fill] (9,3.2) node[right]{\tiny{$r_{1}$}};
\draw[fill] (9,2.2) node[right]{\tiny{$r$}};
\draw[fill] (10.5,1.8) node[right]{\tiny{$r_{2}$}};
\draw[fill] (-5,3.2) node[right]{\tiny{$r_{1}$}};
\draw[fill] (-5,2.2) node[right]{\tiny{$r$}};
\draw[fill] (-3.5,1.8) node[right]{\tiny{$r_{2}$}};
\draw[fill] (4.5,1) node[right]{\tiny{$\lambda=\sqrt{r_{1}r_{2}}$}};
\draw[fill] (7,1) node[right]{\tiny{$\lambda=-\sqrt{r_{1}r_{2}}$}};
\draw[fill] (4.5,-1) node[right]{\tiny{$\lambda=\sqrt{r_{1}r_{2}}$}};
\draw[fill] (7,-1) node[right]{\tiny{$\lambda=-\sqrt{r_{1}r_{2}}$}};
\draw[fill] (7,4) node[right]{\tiny{$\lambda=-i\sqrt{r_{1}r_{2}}$}};
\draw[fill] (7,-4) node[right]{\tiny{$\lambda=i\sqrt{r_{1}r_{2}}$}};
\draw[fill] (11.5,0) node[above]{\tiny{$\lambda=-\sqrt{r_{1}r_{2}}$}};
\draw[fill] (2.3,0) node[above]{\tiny{$\lambda=\sqrt{r_{1}r_{2}}$}};
\draw[fill] (-8.5,1) node{\tiny{$\lambda=-\sqrt{r_{1}r_{2}}$}};
\draw[fill] (-8.5,-1) node{\tiny{$\lambda=-\sqrt{r_{1}r_{2}}$}};
\draw[fill] (-7,1) node[right]{\tiny{$\lambda=\sqrt{r_{1}r_{2}}$}};
\draw[fill] (-7,-1) node[right]{\tiny{$\lambda=\sqrt{r_{1}r_{2}}$}};
\draw[fill] (-7,4) node[right]{\tiny{$\lambda=i\sqrt{r_{1}r_{2}}$}};
\draw[fill] (-7,-4) node[right]{\tiny{$\lambda=-i\sqrt{r_{1}r_{2}}$}};
\draw[fill] (-11.5,0) node[above]{\tiny{$\lambda=-\sqrt{r_{1}r_{2}}$}};
\draw[fill] (-2.3,0) node[above]{\tiny{$\lambda=\sqrt{r_{1}r_{2}}$}};
\draw[fill] (-11,-3) node{\tiny{$S_{1}$}};
\draw[fill] (3,-3) node{\tiny{$S_{2}$}};
\end{tikzpicture}}
\noindent {\small \textbf{Figure 1.} The branch cut on the two-sheeted Riemann surface defined  by $\lambda=\sqrt{k^{2}+q_{0}^{2}}$, where on the sheet $S_{1}:$ $\lambda=\sqrt{r_{1}r_{2}}e^{i(\theta_{1}+\theta_{2})/2}$ and on the sheet $S_{2}:$ $\lambda=-\sqrt{r_{1}r_{2}}e^{i(\theta_{1}+\theta_{2})/2}$ with $-\pi/2<\theta_{1},\theta_{2}<3\pi/2$.}

Further the following transformation is  introduced
\begin{align}\label{Q5}
z=\lambda(k)+k,
\end{align}
which allow us to work in the $z$-plane instead of the two-sheeted Riemann surface. In addition, two single-valued functions are derived
\begin{align}\label{Q6}
k(z)=\frac{1}{2}\left(z-\frac{q_{0}^{2}}{z}\right),\quad \lambda(z)=\frac{1}{2}\left(z+\frac{q_{0}^{2}}{z}\right).
\end{align}

Using the eigenvalues $\pm i\lambda$ of the matrix $X_{\pm}$, the eigenvectors are expressed as
\begin{align}\label{Q7}
Y_{\pm}(z)=I+\frac{i}{z}\sigma_{3}Q_{\pm},
\end{align}
then  matrices $X_{\pm}$ and $T_{\pm}$ can be diagonalized
\begin{align}\label{Q8}
X_{\pm}=Y_{\pm}(i\lambda\sigma_{3})Y_{\pm}^{-1},~~
T_{\pm}=Y_{\pm}(-2ik\lambda\sigma_{3})Y_{\pm}^{-1}.
\end{align}
Note that the continuous spectrum $\Sigma_{k}$ is composed of all $k$ satisfying $\lambda(k)\in R$, namely $\Sigma_{k}=R\cup i[-q_{0},q_{0}]$, which corresponds to $\Sigma_{z}=R\cup C_{0}$ in the $z$-plane, where $C_{0}=\{|z|=q_{0}\}$ in the complex $z$-plane. For convenience, taking $\Sigma=\Sigma_{z}$, for all $z\in\Sigma$, we define the Jost functions $\psi_{\pm}$  as the solutions of the Lax pair \eqref{Q3}
\begin{align}\label{Q9}
\phi_{\pm}(x,t;z)= Y_{\pm}(z)e^{i\theta(x,t;z)\sigma_{3}}+o(1) \quad x\rightarrow \pm\infty,
\end{align}
with $\theta=\lambda(z)\left(x-2k(z)t\right)$.  Further the modified eigenfunctions are introduced to eliminate exponential oscillations
\begin{align}\label{Q10}
u_{\pm}(x,t;z)=\phi_{\pm}(x,t;z)e^{-i\theta(x,t;z)\sigma_{3}}\rightarrow Y_{\pm}(z),~ x\rightarrow\pm\infty,
\end{align}
then the $u_{\pm}(x,t;z)$ are expressed as the following integral equations as usual
\begin{align}\label{Q11}
\begin{split}
u_{-}(x,t;z)=Y_{-}+\int_{-\infty}^{x}Y_{-}e^{i\lambda(x-y)\sigma_{3}}Y_{-}^{-1}\Delta Q_{-}(y,t)u_{-}(y,t;z)e^{-i\lambda(x-y)\sigma_{3}}\, dy,\\
u_{+}(x,t;z)=Y_{+}-\int_{x}^{\infty}Y_{+}e^{i\lambda(x-y)\sigma_{3}}Y_{+}^{-1}\Delta Q_{+}(y,t)u_{+}(y,t;z)e^{-i\lambda(x-y)\sigma_{3}}\, dy.
\end{split}
\end{align}
The analytic properties of the eigenfunctions $u_{\pm}$ have been proved in detail in \cite{Biondini-JMP2014}. Here we rephrase it as the following proposition.
\begin{prop}\label{p1}
If $q(x,t)-q_{\pm}\in L^{1}(R)$, the modified eigenfunctions $u_{\pm}(x,t;z)$ can be analytically extended onto the corresponding regions of the $z$-plane, that is
\begin{align}\label{Q12}
u_{+,1}(x,t;z),~ u_{-,2}(x,;z)~\text{are analytic in}~ D^{+}=\left\{z\in C:(|z|^{2}-q_{0}^{2})Im~z>0\right\},\\
u_{+,2}(x,t;z),~ u_{-,1}(x,;z)~ \text{are analytic in}~ D^{-}=\left\{z\in C:(|z|^{2}-q_{0}^{2})Im~z<0\right\},
\end{align}
where $u_{\pm,j}(x,t;z)$ $(j=1,2)$ represent the $j$-th column of the eigenfunctions $u_{\pm}(x,t;z)$.
\end{prop}
\subsection{Scattering matrix}
Using the fact $\text{tr}X=\text{tr}T=0$, combined with Abel formula, we have $(\det\phi)_{x}=(\det\phi)_{t}=0$ and $\det(u_{\pm})=\det(\phi_{\pm})$. Thus one has
\begin{align}\label{Q13}
\det(u_{\pm})=\lim_{x\rightarrow\pm\infty}\det(u_{\pm})=\det(Y_{\pm})=1+q_{0}^{2}/z^{2}\triangleq \varpi(z)\neq0,~z\in\Sigma_{0},
\end{align}
where $\Sigma_{0}=\Sigma\setminus\{\pm iq_{0}\}$. Note that $\phi_{\pm}$ are the fundamental matrix solutions of the Lax pair \eqref{Q3}, there exists a matrix $S(z)$ independent of the independent variables $x,t$ that satisfies
\begin{align}\label{Q14}
\phi_{+}(x,t;z)=\phi_{-}(x,t;z)S(z),~~z\in\Sigma_{0},
\end{align}
which is called the scattering matrix with $S(z)=s_{ij}(z)$ ($i,j=1,2$). By directly calculating equation \eqref{Q14}, the elements of matrix $S(z)$ can be expressed as
\addtocounter{equation}{1}
\begin{align}
s_{11}(z)\varpi=Wr\left(\phi_{+,1},\phi_{-,2}\right), \quad
s_{22}(z)\varpi=Wr\left(\phi_{-,1},\phi_{+,2}\right),\tag{\theequation a} \label{Q15}\\
s_{12}(z)\varpi=Wr\left(\phi_{+,2},\phi_{-,2}\right),\quad
s_{21}(z)\varpi=Wr\left(\phi_{-,1},\phi_{+,1}\right).\tag{\theequation b}\label{Q16}
\end{align}
It follows that from the proposition \ref{p1}, under the condition $q-q_{\pm}\in L^{1}(R^{\pm})$ the scattering data $s_{11}(z)$ and $s_{22}(z)$ defined by \eqref{Q15} are analytic in $D^{+}$ and $D^{-}$, as well as they can be analytically extend to $D^{+}\cup\Sigma_{0}$ and $D^{-}\cup\Sigma_{0}$, respectively. In addition, the scattering data $s_{12}(z)$ and $s_{21}(z)$ defined by \eqref{Q16} are continuous in $\Sigma_{0}$. Finally, the reflection coefficients appeared in the RH problem are given as
\begin{align}\label{Q17}
\rho(z)=\frac{s_{21}(z)}{s_{11}(z)},~~
\widetilde{\rho}(z)=\frac{s_{12}(z)}{s_{22}(z)}.
\end{align}

\subsection{Symmetry properties}
It is noticed that there are two independent symmetries in the scattering problem due to the introduction of Riemann surface.  One of them is that $z\mapsto z^{*}$ implies $(k,\lambda)\mapsto(k^{*},\lambda^{*})$ and the other is $z\mapsto -q_{0}^{2}/z$ implies $(k,\lambda)\mapsto(k,-\lambda)$.
\begin{prop}\label{p2}
The Jost functions $\phi_{\pm}$ and the scattering matrix $S(z)$ satisfy the following symmetry
\begin{enumerate}[(1)]
\item First symmetry
\begin{align}\label{Q18}
\phi_{\pm}(z)=-\sigma\phi_{\pm}^{*}(z^{*})\sigma,~~S^{*}(z^{*})=-\sigma S(z)\sigma, ~~z\in\Sigma.
\end{align}

\item Second symmetry
\begin{align}\label{Q19}
\phi_{\pm}(z)=\frac{1}{z}\phi_{\pm}(-q^{2}_{0}/z)\sigma_{3} Q_{\pm},~~S(-q^{2}_{0}/z)=\sigma_{3}Q_{-} S(z)(\sigma_{3}Q_{+})^{-1}, ~~z\in\Sigma.
\end{align}
\end{enumerate}
\end{prop}

Using the relationship between reflection coefficients and scattering data \eqref{Q17}, we can get the symmetry of reflection coefficients, which plays an important role in making Laurent series.
\begin{cor}\label{p3}
The reflection coefficients admit that
\begin{align}\label{Q20}
\rho(z)=-\widetilde{\rho}^{*}(z^{*})=\frac{q_{-}}{q_{-}^{*}}\widetilde{\rho}(-q^{2}_{0}/z)=
-\frac{q_{-}^{*}}{q_{-}}\rho^{*}(-q^{2}_{0}/z^{*}),~~~z\in\Sigma.
\end{align}
\end{cor}
\subsection{Asymptotic behaviors}
Similar to symmetry, due to the introduction of Riemann surface, the modified eigenfunctions $u_{\pm}$ and scattering matrix $S(z)$ correspond to two cases in the $z$-plane when the spectral parameter $k$ tend to infinity, i.e., $z\rightarrow0$ and $z\rightarrow\infty$ in $z$-plane.
\begin{prop}\label{p4}
The modified eigenfunctions $u_{\pm}$ and the scattering matrix $S(z)$ obey the following asymptotic behaviors, which have been proved in \cite{Biondini-JMP2014}
\begin{align}\label{Q21}
\begin{split}
&u_{\pm}(z)=I+O(1/z),~~~z\rightarrow\infty,\\
&u_{\pm}(z)=\frac{i}{z}\sigma_{3}Q_{\pm}+O(1),~~~z\rightarrow0,\\
&S(z)=I+O(1/z),~~~z\rightarrow\infty,\\
&S(z)=\text{diag}(q_{-}/q_{+},q_{+}/q_{-})+O(z),~~~z\rightarrow0.
\end{split}
\end{align}
\end{prop}
\subsection{Discrete spectrum}
In general, the set of discrete spectrum is composed of all $z$ that make the eigenfunction exist in $L^{2}(R)$. In addition, Biondini and Kova\u{c}i\u{c} have shown that the set is the zeros of scattering data $s_{11}(z)$ and $s_{22}(z)$ in $D^{+}$ and $D^{-}$ in \cite{Biondini-JMP2014}, respectively. Suppose that $s_{11}(z)$ has  finite zeros in $D^{+}\cap\{z\in C:Im~z>0\}$ denoted by $\{z_{n}\}_{n=1}^{N}$, then employing the relationships \eqref{Q18} and  \eqref{Q19} can derive that
\begin{align}\label{Q22}
s_{11}(z_{n})=0\Longleftrightarrow s_{22}(z_{n}^{*})=0 \Longleftrightarrow s_{22}(-q_{0}^{2}/z_{n})=0\Longleftrightarrow
s_{11}(-q_{0}^{2}/z_{n}^{*})=0,
\end{align}
thus the discrete spectrum set is composed of $Z=\{z_{n},z_{n}^{*},-q_{0}^{2}/z_{n},-q_{0}^{2}/z_{n}^{*}\}$, which can be shown in figure 2.

\centerline{\begin{tikzpicture}[scale=0.65]
\draw[-][thick](4,4)--(4,-4);
\draw[-][thick](4,-4)--(-4,-4);
\draw[-][thick](-4,-4)--(-4,4);
\draw[-][thick](-4,4)--(4,4);
\draw[fill] (0,-2)node[below]{} circle [radius=0.055];
\draw[fill] (0,2)node[below]{} circle [radius=0.055];
\draw[fill] (2,0)node[below]{} circle [radius=0.055];
\draw[fill] (-2,0)node[below]{} circle [radius=0.055];
\draw[fill] (0,0)node[below]{} circle [radius=0.055];
\path [fill=white] (-4,0)--(-4,4) to (4,4) -- (4,0);
\filldraw[gray, line width=0.5](-4,0)--(-2,0) arc (180:0:2);
\path [fill=gray] (-4,0)--(-4,-4) to (4,-4) -- (4,0);
\filldraw[white, line width=0.5](-4,0)--(-2,0) arc (-180:0:2);
\draw[->][thick](-4,0)--(-3,0);
\draw[-][thick](-3,0)--(-2,0)node[below right]{\footnotesize$0^{-}$};;
\draw[-][thick](-2,0)--(-1,0);
\draw[<-][thick](-1,0)--(0,0);
\draw[-][thick](0,0)--(1,0);
\draw[<-][thick](1,0)--(2,0)node[below right]{\footnotesize$0^{+}$};
\draw[->][thick](2,0)--(3,0);
\draw[->][thick](3,0)--(4,0)[thick]node[right]{$Rez$};
\draw[-][thick](0,0)--(0,1);
\draw[-][thick](0,1)--(0,2)node[below right]{\footnotesize$iq_{0}$};
\draw[-][thick](0,2)--(0,3);
\draw[->][thick](0,3)--(0,4)[thick]node[right]{$Imz$};
\draw[-][thick](0,0)--(0,-1);
\draw[-][thick](0,-1)--(0,-2)node[below right]{\footnotesize$-iq_{0}$};
\draw[-][thick](0,-2)--(0,-3);
\draw[-][thick](0,-3)--(0,-4);
\draw[->][red, line width=0.8] (2,0) arc(0:220:2);
\draw[->][red, line width=0.8] (2,0) arc(0:330:2);
\draw[->][red, line width=0.8] (2,0) arc(0:-330:2);
\draw[->][red, line width=0.8] (2,0) arc(0:-220:2);
\draw[fill][blue] (2.5,2.5) circle [radius=0.035][thick]node[right]{\footnotesize$z_{n}$};
\draw[fill] (2.5,-2.5) circle [radius=0.035][thick]node[right]{\footnotesize$z_{n}^{*}$};
\draw[fill] (-1,1) circle [radius=0.035][thick]node[right]{\footnotesize$-\frac{q_{0}^{2}}{z_{n}}$};
\draw[fill][blue] (-1,-1) circle [radius=0.035][thick]node[right]{\footnotesize$-\frac{q_{0}^{2}}{z_{n}^{*}}$};
\end{tikzpicture}}

\centerline{\noindent {\small \textbf{Figure 2.} The distribution of the set of discrete spectral points on the $z$-plane.} }

\subsection{Riemann-Hilbert problem}
In order to establish a suitable Riemann-Hilbert problem, a piecewise analytic function $M(x,t;z)$ is introduced as usual
\begin{align}\label{Q23}
M(x,t;z)=\left\{\begin{aligned}
\left(u_{-,1}(x,t;z),\frac{u_{+,2}(x,t;z)}{s_{22}(z)}\right), \quad z\in D^{-},\\
\left(\frac{u_{+,1}(x,t;z)}{s_{11}(z)},u_{-,2}(x,t;z)\right), \quad z\in D^{+}.
\end{aligned}\right.
\end{align}
\begin{prop}\label{p5}
Find a matrix $M(z)=M(x,t;z)$ satisfies the following conditions
\begin{enumerate}[(1)]
\item Analyticity: $M(z)$ is a meromorphic function in $C\setminus\Sigma$.

\item Jump condition:
\begin{align}\label{Q24}
M^{-}(z)=M^{+}(z)G(z), ~~z\in\Sigma,
\end{align}
where the jump matrix $G(z)$ is defined by
\begin{align*}
G(x,t;z)=\left(
\begin{array}{ccc}
  1 & \widetilde{\rho}(z)e^{2i\theta(z)} \\
  -\rho(z)e^{-2i\theta(z)} & 1-\rho(z)\widetilde{\rho}(z)
\end{array} \right). \notag
\end{align*}
\item Asymptotic condition:
\begin{align}\label{Q25}
\begin{split}
&M^{\pm}(z)=I+O(1/z),~~~z\rightarrow\infty,\\
&M^{\pm}(z)=\frac{i}{z}\sigma_{3}Q_{-}+O(1),~~~z\rightarrow0.
\end{split}
\end{align}
\item The relationship between the solutions of NLS equation and RH problem:
\begin{align}\label{Q26}
q(x,t)=-i\lim_{z\rightarrow\infty}\left(zM(z)\right)_{12}.
\end{align}
\end{enumerate}
\end{prop}
\section{The solution of the NLS with a single high-order pole}

Taking a single high-order pole into account firstly, i.e., $z_{1}\in D^{+}$  is the $N$-order pole of the scattering data $s_{11}(z)$, one obtains another $N$-order pole $-q_{0}^{2}/z_{1}^{*}\in D^{+}$ from the symmetry \eqref{Q22}. Additionally, $z_{1}^{*}$,
$-q_{0}^{2}/z_{1}\in D^{-}$ are the $N$-order pole of the scattering data $s_{22}(z)$. Taking $v_{1}=z_{1}$ and $v_{2}=-q_{0}^{2}/z_{1}^{*}$, then $s_{11}(z)$ can be expressed as
\begin{align}\label{Q27}
s_{11}(z)=(z-v_{1})^{N}(z-v_{2})^{N}s_{11}^{(0)}(z),
\end{align}
where $s_{11}^{(0)}(z)\neq0$ for all $z\in D^{+}$. In accordance with  Laurent series expansion, the reflection coefficients $\rho(z)$ and $\widetilde{\rho}^{*}(z^{*})$ can be determined by
\begin{align}\label{Q28}
&\rho(z)=\rho_{0}(z)+\sum_{j=1}^{N}\frac{\rho_{1,j}}{(z-v_{1})^{j}},\quad\quad~~
\rho(z)=\widehat{\rho}_{0}(z)+\sum_{j=1}^{N}\frac{\rho_{2,j}}{(z-v_{2})^{j}},\\
&\rho^{*}(z^{*})=\rho^{*}_{0}(z^{*})+\sum_{j=1}^{N}\frac{\rho_{1,j}^{*}}{(z-v_{1}^{*})^{j}},\quad
\rho^{*}(z^{*})=\widehat{\rho}^{*}_{0}(z^{*})+\sum_{j=1}^{N}\frac{\rho_{2,j}^{*}}{(z-v_{2}^{*})^{j}},
\end{align}
where $\rho_{m,j}$ ($j=1,2$ and $m=1,2,\ldots,N$)
\begin{align*}
\rho_{m,j}=\lim_{z\rightarrow v_{m}}\frac{1}{(N-j)!}\frac{\partial^{N-j}}{\partial z^{N-j}}
\left[(z-v_{m})^{N}\rho_{m}(z)\right],
\end{align*}
and $\rho_{0}(z)$, $\widehat{\rho}_{0}(z)$ are analytic functions. Based on the definition of $M(z)$ defined by \eqref{Q23}, it follows from the asymptotic condition shown in proposition \ref{p4} that $M(z)$ can be expanded at the poles as follows
\begin{align}
&M_{11}(x,t;z)=1+\sum_{s=1}^{N}\left(\frac{F_{s}(x,t)}{(z-v_{1})^{s}}+
\frac{H_{s}(x,t)}{(z-v_{2})^{s}}\right),\label{Q29-a}\\
&M_{12}(x,t;z)=\frac{i}{z}q_{-}+\sum_{s=1}^{N}\left(\frac{G_{s}(x,t)}{(z-v_{1}^{*})^{s}}+
\frac{L_{s}(x,t)}{(z-v_{2}^{*})^{s}}\right),\label{Q29-b}
\end{align}
where $F_{s}(x,t)$, $H_{s}(x,t)$, $G_{s}(x,t)$ and $L_{s}(x,t)$ ($s=1,2,\ldots,N$) are unknown functions, which can be determined by the coefficients of Taylor expansion of exponential functions $e^{\pm2i\theta(z)}$ and function $M(z)$ at the poles $\{v_{j},v_{j}^{*}\}_{j=1}^{2}$.  Note that once these functions are determined, the solution of the focusing NLS equation can also be obtained according to the relation \eqref{Q26}.

In what follows, taking the Taylor expansion into account, one has
\addtocounter{equation}{1}
\begin{align}
e^{-2i\theta(z)}&=\sum_{\ell=0}^{+\infty}f_{1,\ell}(x,t)(z-v_{1})^{\ell},\quad
e^{-2i\theta(z)}=\sum_{\ell=0}^{+\infty}f_{2,\ell}(x,t)(z-v_{2})^{\ell},\tag{\theequation a}\label{Q30}\\
e^{2i\theta(z)}&=\sum_{\ell=0}^{+\infty}f^{*}_{1,\ell}(x,t)(z-v_{1}^{*})^{\ell},\quad
e^{2i\theta(z)}=\sum_{\ell=0}^{+\infty}f^{*}_{2,\ell}(x,t)(z-v_{2}^{*})^{\ell},\tag{\theequation b}\label{Q31}\\
M_{11}(x,t;z)&=\sum_{\ell=0}^{+\infty}\mu_{1,\ell}(x,t)(z-v_{1}^{*})^{\ell},\quad
M_{11}(x,t;z)=\sum_{\ell=0}^{+\infty}\mu_{2,\ell}(x,t)(z-v_{2}^{*})^{\ell},\tag{\theequation c}\label{Q32}\\
M_{12}(x,t;z)&=\sum_{\ell=0}^{+\infty}\zeta_{1,\ell}(x,t)(z-v_{1})^{\ell},\quad
M_{12}(x,t;z)=\sum_{\ell=0}^{+\infty}\zeta_{2,\ell}(x,t)(z-v_{2})^{\ell},\tag{\theequation d}\label{Q33}
\end{align}
where
\addtocounter{equation}{1}\label{Xishu}
\begin{align}
f_{j,\ell}(x,t)&=\lim_{z\rightarrow v_{j}}\frac{1}{\ell !}\frac{\partial^{\ell}}{\partial z^{\ell}}
e^{-2i\theta(z)},\tag{\theequation a}\label{Q34}\\
\mu_{j,\ell}(x,t)&=\lim_{z\rightarrow v_{j}}\frac{1}{\ell !}\frac{\partial^{\ell}}
{\partial z^{\ell}}M_{11}(x,t;z),\tag{\theequation b}\label{Q35}\\
\nu_{j,\ell}(x,t)&=\lim_{z\rightarrow v_{j}}\frac{1}{\ell !}\frac{\partial^{\ell}}
{\partial z^{\ell}}M_{12}(x,t;z),\quad \ell=0,1,2,\cdots, ~j=1,2.\tag{\theequation c}\label{Q36}
\end{align}

Let $z\in D^{+}$, the function $M(z)$ has the expansions at $z=v_{1}$
\begin{align}
&M_{12}(z)=u_{-,12}(z)=\sum_{\ell=0}^{\infty}\zeta_{1,\ell}(z-v_{1})^{\ell},\label{Q37}\\
&M_{11}(z)=\frac{u_{+,12}(z)}{s_{11}(z)}=u_{-,11}(z)+\rho(z)e^{-2i\theta(z)}
u_{-,12}(z).\label{Q38}
\end{align}
Substituting equations \eqref{Q29-a} and \eqref{Q37} into \eqref{Q38} yields
\begin{align}\label{Q39}
F_{s}(x,t)=\sum_{j=s}^{N}\sum_{\ell=0}^{j-s}\rho_{1,j}f_{1,j-s-\ell}
(x,t)\zeta_{1,\ell}(x,t).
\end{align}
Then taking the expansions at $z=v_{2}\in D^{+}$ yields
\begin{align}\label{Q40}
H_{s}(x,t)=\sum_{j=s}^{N}\sum_{\ell=0}^{j-s}\rho_{2,j}f_{2,j-s-\ell}
(x,t)\zeta_{2,\ell}(x,t).
\end{align}
Let $z\in D^{-}$, the expressions of $G_{s}(x,t)$ and $L_{s}(x,t)$ can be derived in the same way
\begin{align}\label{Q41}
\begin{split}
G_{s}(x,t)=-\sum_{j=s}^{N}\sum_{\ell=0}^{j-s}\rho_{1,j}^{*}f_{1,j-s-\ell}^{*}
(x,t)\mu_{1,\ell}(x,t),\\
L_{s}(x,t)=-\sum_{j=s}^{N}\sum_{\ell=0}^{j-s}\rho_{2,j}^{*}f_{2,j-s-\ell}^{*}
(x,t)\mu_{2,\ell}(x,t).
\end{split}
\end{align}

Notice that $\mu_{j,\ell}$ and $\zeta_{j,\ell}$ ($j=1,2$) can be represented by these four functions $F_{s}(x,t)$, $H_{s}(x,t)$, $G_{s}(x,t)$ and $L_{s}(x,t)$ ($s=1,2,\ldots,N$). Specifically speaking, in accordance with these expressions defined by \eqref{Q34}-\eqref{Q36}, \eqref{Q29-a} and \eqref{Q29-b}, we obtain
\begin{align*}
\zeta_{1,\ell}(x,t)=\frac{(-1)^{\ell}}{(v_{1})^{\ell+1}}iq_{-}+\sum_{s=1}^{N}\left(
                                      \begin{array}{c}
                                        s+\ell-1 \\
                                        \ell \\
                                      \end{array}
                                    \right)
\left\{\frac{(-1)^{\ell}G_{s}(x,t)}{(v_{1}-v_{1}^{*})^{\ell+s}}+
\frac{(-1)^{\ell}L_{s}(x,t)}{(v_{1}-v_{2}^{*})^{\ell+s}}\right\},\quad \ell=0,1,2,\cdots, \\
\zeta_{2,\ell}(x,t)=\frac{(-1)^{\ell}}{(v_{2})^{\ell+1}}iq_{-}+\sum_{s=1}^{N}\left(
                                      \begin{array}{c}
                                        s+\ell-1 \\
                                        \ell \\
                                      \end{array}
                                    \right)
\left\{\frac{(-1)^{\ell}G_{s}(x,t)}{(v_{2}-v_{1}^{*})^{\ell+s}}+
\frac{(-1)^{\ell}L_{s}(x,t)}{(v_{2}-v_{2}^{*})^{\ell+s}}\right\},\quad \ell=0,1,2,\cdots, \\
\mu_{1,\ell}(x,t)=\left\{\begin{aligned}
&1+\sum_{s=1}^{N}\left(\frac{(-1)^{\ell}F_{s}(x,t)}{(v_{1}^{*}-v_{1})^{s+\ell}}+
\frac{(-1)^{\ell}H_{s}(x,t)}{(v_{1}^{*}-v_{2})^{\ell+s}}\right), \qquad\qquad \quad \ell=0,\\
&\sum_{s=1}^{N}\left(
                                      \begin{array}{c}
                                        s+\ell-1 \\
                                        \ell \\
                                      \end{array}
                                    \right)
\left\{\frac{(-1)^{\ell}F_{s}(x,t)}{(v_{1}^{*}-v_{1})^{\ell+s}}+
\frac{(-1)^{\ell}H_{s}(x,t)}{(v_{1}^{*}-v_{2})^{\ell+s}}\right\},\quad \ell=1,2,\cdots.\\
\end{aligned}\right. \\
\mu_{2,\ell}(x,t)=\left\{\begin{aligned}
&1+\sum_{s=1}^{N}\left(\frac{(-1)^{\ell}F_{s}(x,t)}{(v_{2}^{*}-v_{1})^{s+\ell}}+
\frac{(-1)^{\ell}H_{s}(x,t)}{(v_{2}^{*}-v_{2})^{\ell+s}}\right), \qquad\qquad \quad \ell=0,\\
&\sum_{s=1}^{N}\left(
                                      \begin{array}{c}
                                        s+\ell-1 \\
                                        \ell \\
                                      \end{array}
                                    \right)
\left\{\frac{(-1)^{\ell}F_{s}(x,t)}{(v_{2}^{*}-v_{1})^{\ell+s}}+
\frac{(-1)^{\ell}H_{s}(x,t)}{(v_{2}^{*}-v_{2})^{\ell+s}}\right\},\quad \ell=1,2,\cdots.\\
\end{aligned}\right.
\end{align*}
Substituting the above equations into \eqref{Q39}, \eqref{Q40} and \eqref{Q41}, one has the following closed algebraic system
\begin{align}
F_{s}(x,t)=iq_{-}&\sum_{j=s}^{N}\sum_{\ell=0}^{j-s}\frac{(-1)^{\ell}}{(v_{1})^{\ell+1}}\rho_{1,j}
f_{1,j-\ell-s}(x,t)+\notag\\
&\sum_{j=s}^{N}\sum_{\ell=0}^{j-s}\sum_{p=1}^{N}\left(
                                      \begin{array}{c}
                                        p+\ell-1 \\
                                        \ell \\
                                      \end{array}
                                    \right)\rho_{1,j}
f_{1,j-\ell-s}(x,t)
\left\{\frac{(-1)^{\ell}G_{p}(x,t)}{(v_{1}-v_{1}^{*})^{\ell+p}}+
\frac{(-1)^{\ell}L_{p}(x,t)}{(v_{1}-v_{2}^{*})^{\ell+p}}\right\},\label{Q42}
\end{align}
\begin{align}
H_{s}(x,t)=iq_{-}&\sum_{j=s}^{N}\sum_{\ell=0}^{j-s}\frac{(-1)^{\ell}}{(v_{2})^{\ell+1}}\rho_{2,j}
f_{2,j-\ell-s}(x,t)+\notag\\
&\sum_{j=s}^{N}\sum_{\ell=0}^{j-s}\sum_{p=1}^{N}\left(
                                      \begin{array}{c}
                                        p+\ell-1 \\
                                        \ell \\
                                      \end{array}
                                    \right)\rho_{2,j}
f_{2,j-\ell-s}
\left\{\frac{(-1)^{\ell}G_{p}(x,t)}{(v_{2}-v_{1}^{*})^{\ell+p}}+
\frac{(-1)^{\ell}L_{p}(x,t)}{(v_{2}-v_{2}^{*})^{\ell+p}}\right\},\label{Q43}\\
G_{s}(x,t)=-&\sum_{j=s}^{N}  \rho_{1,j}^{*}f_{1,j-s}^{*}-\notag\\
&\sum_{j=s}^{N}\sum_{\ell=0}^{j-s}\sum_{p=1}^{N}\left(
                                      \begin{array}{c}
                                        p+\ell-1 \\
                                        \ell \\
                                      \end{array}
                                    \right)\rho_{1,j}^{*}
f_{1,j-\ell-s}^{*}
\left\{\frac{(-1)^{\ell}F_{p}(x,t)}{(v_{1}^{*}-v_{1})^{\ell+p}}+
\frac{(-1)^{\ell}H_{p}(x,t)}{(v_{1}^{*}-v_{2})^{\ell+p}}\right\},\label{Q44}\\
L_{s}(x,t)=-&\sum_{j=s}^{N}  \rho_{2,j}^{*}f_{2,j-s}^{*}-\notag\\
&\sum_{j=s}^{N}\sum_{\ell=0}^{j-s}\sum_{p=1}^{N}\left(
                                      \begin{array}{c}
                                        p+\ell-1 \\
                                        \ell \\
                                      \end{array}
                                    \right)\rho_{2,j}^{*}
f_{2,j-\ell-s}^{*}
\left\{\frac{(-1)^{\ell}F_{p}(x,t)}{(v_{2}^{*}-v_{1})^{\ell+p}}+
\frac{(-1)^{\ell}H_{p}(x,t)}{(v_{2}^{*}-v_{2})^{\ell+p}}\right\}.\label{Q45}
\end{align}

For convenience, taking the notations
\begin{align}
&|\eta_{1}\rangle=(\eta_{1,1},\eta_{1,2},\ldots,\eta_{1,N})^{T},~~\eta_{1,s}=iq_{-}
\sum_{j=s}^{N}\sum_{\ell=0}^{j-s}\frac{(-1)^{\ell}}{(v_{1})^{\ell+1}}\rho_{1,j}
f_{1,j-\ell-s}(x,t),\label{xx1}\\
&|\eta_{2}\rangle=(\eta_{2,1},\eta_{2,2},\ldots,\eta_{2,N})^{T},~~\eta_{2,s}=iq_{-}
\sum_{j=s}^{N}\sum_{\ell=0}^{j-s}\frac{(-1)^{\ell}}{(v_{2})^{\ell+1}}\rho_{2,j}
f_{2,j-\ell-s}(x,t),\\
&|\widehat{\eta}_{1}\rangle=(\tilde{\eta}_{1,1},\tilde{\eta}_{1,2},\ldots,\tilde{\eta}_{1,N})^{T},
~~~~~~~~~~\tilde{\eta}_{1,s}=-\sum_{j=s}^{N}\rho_{1,j}^{*}f_{1,j-s}^{*},\\
&|\widehat{\eta}_{2}\rangle=(\tilde{\eta}_{2,1},\tilde{\eta}_{2,2},\ldots,\tilde{\eta}_{2,N})^{T},
~~~~~~~~~~\tilde{\eta}_{2,s}=-\sum_{j=s}^{N}\rho_{2,j}^{*}f_{2,j-s}^{*},\\
&\Theta_{1}=[\Theta_{1,sp}]_{N\times N}=\sum_{j=s}^{N}\sum_{\ell=0}^{j-s}\left(
                                      \begin{array}{c}
                                        p+\ell-1 \\
                                        \ell \\
                                      \end{array}
                                    \right)
\frac{(-1)^{\ell}\rho_{1,j}f_{1,j-\ell-s}(x,t)}{(v_{1}-v_{1}^{*})^{\ell+p}},\\
&\Theta_{2}=[\Theta_{2,sp}]_{N\times N}=\sum_{j=s}^{N}\sum_{\ell=0}^{j-s}\left(
                                      \begin{array}{c}
                                        p+\ell-1 \\
                                        \ell \\
                                      \end{array}
                                    \right)
\frac{(-1)^{\ell}\rho_{1,j}f_{1,j-\ell-s}(x,t)}{(v_{1}-v_{2}^{*})^{\ell+p}},\\
&\Theta_{3}=[\Theta_{3,sp}]_{N\times N}=\sum_{j=s}^{N}\sum_{\ell=0}^{j-s}\left(
                                      \begin{array}{c}
                                        p+\ell-1 \\
                                        \ell \\
                                      \end{array}
                                    \right)
\frac{(-1)^{\ell}\rho_{2,j}f_{2,j-\ell-s}(x,t)}{(v_{2}-v_{1}^{*})^{\ell+p}},\\
&\Theta_{4}=[\Theta_{4,sp}]_{N\times N}=\sum_{j=s}^{N}\sum_{\ell=0}^{j-s}\left(
                                      \begin{array}{c}
                                        p+\ell-1 \\
                                        \ell \\
                                      \end{array}
                                    \right)
\frac{(-1)^{\ell}\rho_{2,j}f_{2,j-\ell-s}(x,t)}{(v_{2}-v_{2}^{*})^{\ell+p}},\\
&|F\rangle=(F_{1},F_{2},\ldots,F_{N})^{T},~~~|H\rangle=(H_{1},H_{2},\ldots,H_{N})^{T},\\
&|G\rangle=(G_{1},G_{2},\ldots,G_{N})^{T},~~~|L\rangle=(L_{1},L_{2},\ldots,L_{N})^{T},\label{xx2}
\end{align}
the closed algebraic system \eqref{Q42}-\eqref{Q45} then can be written as the matrix form
\begin{align}\label{1}
\begin{split}
I|F\rangle+\mathbf{0}|H\rangle-\Theta_{1}|G\rangle-\Theta_{2}|L\rangle=|\eta_{1}\rangle,\\
\mathbf{0}|F\rangle+I|H\rangle-\Theta_{3}|G\rangle-\Theta_{4}|L\rangle=|\eta_{2}\rangle,\\
\Theta_{1}^{*}|F\rangle+\Theta_{2}^{*}|H\rangle+I|G\rangle+\mathbf{0}|L\rangle=
|\widehat{\eta}_{1}\rangle,\\
\Theta_{3}^{*}|F\rangle+\Theta_{4}^{*}|H\rangle+\mathbf{0}|G\rangle+I|L\rangle=
|\widehat{\eta}_{2}\rangle.
\end{split}
\end{align}

Further taking
$
\Theta=\left(
         \begin{array}{cc}
           \Theta_{1} & \Theta_{2} \\
           \Theta_{3} & \Theta_{4} \\
         \end{array}
       \right),~~|K_{1}\rangle =(|\eta_{1}\rangle,|\eta_{2}\rangle)^{T},~~
  |K_{2}\rangle =(|\widehat{\eta}_{1}\rangle,|\widehat{\eta}_{2}\rangle)^{T},
$
the system \eqref{1} can be solved
\begin{align}
&(|G\rangle,|L\rangle)^{T}=-(I_{\epsilon}+\Theta^{*}\Theta)^{-1}\Theta^{*}
|K_{1}\rangle+(I_{\epsilon}+\Theta^{*}\Theta)^{-1}|K_{2}\rangle,\label{2}\\
&(|F\rangle,|H\rangle)^{T}=-\Theta\left((I_{\epsilon}+\Theta^{*}\Theta)^{-1}\Theta^{*}
|K_{1}\rangle+(I_{\epsilon}+\Theta^{*}\Theta)^{-1}|K_{2}\rangle\right)+|K_{1}\rangle,\label{3}
\end{align}
where $I_{\epsilon}=\left(
                      \begin{array}{cc}
                        I_{N\times N} &  \\
                         & I_{N\times N} \\
                      \end{array}
                    \right)
$.

Applying \eqref{Q29-b} and \eqref{2}, one has
\begin{align}
M_{12}(x,t;z)
=\frac{i}{z}q_{-}+\frac{\det\left(I_{\epsilon}+\Theta^{*}\Theta+|K_{2}\rangle\langle Y_{0}|\right)-\det\left(I_{\epsilon}+\Theta^{*}\Theta+\Theta^{*}|K_{1}\rangle\langle Y_{0}|\right)}{\det\left(I_{\epsilon}+\Theta^{*}\Theta\right)},
\end{align}
where $\langle Y_{0}|=(1,0,\ldots,0,1,0,\ldots,0)_{1\times2N}$.
\begin{thm}
Under the nonzero condition \eqref{Q2}, the solution of focusing NLS with single high-order pole can be derived as the form
\begin{align}\label{Q-jie}
q(x,t)=q_{-}-i\frac{\det\left(I_{\epsilon}+\Theta^{*}\Theta+|K_{2}\rangle\langle Y_{0}|\right)-\det\left(I_{\epsilon}+\Theta^{*}\Theta+\Theta^{*}|K_{1}\rangle\langle Y_{0}|\right)}{\det\left(I_{\epsilon}+\Theta^{*}\Theta\right)}.
\end{align}
\end{thm}

\textbf{Case(A):} For a simple pole, i.e., $N=1$, 
the elements can be derived from \eqref{xx1}-\eqref{xx2}
\begin{align*}
\eta_{1,1}=\frac{iq_{-}\rho_{1,1}f_{1,0}}{v_{1}},~\tilde{\eta}_{1,1}=-r_{1,1}^{*}f_{1,0}^{*},~
\Theta_{1,1}=\frac{\rho_{1,1}f_{1,0}}{v_{1}-v_{1}^{*}},~
\Theta_{1,2}=\frac{\rho_{1,1}f_{1,0}}{v_{1}-v_{2}^{*}},\\
\eta_{2,1}=\frac{iq_{-}\rho_{2,1}f_{2,0}}{v_{2}},~\tilde{\eta}_{2,1}=-r_{2,1}^{*}f_{2,0}^{*},~
\Theta_{2,1}=\frac{\rho_{2,1}f_{2,0}}{v_{2}-v_{1}^{*}},~
\Theta_{2,2}=\frac{\rho_{2,1}f_{2,0}}{v_{2}-v_{2}^{*}}.
\end{align*}

\centerline{{\rotatebox{0}{\includegraphics[width=3.75cm,height=3.5cm,angle=0]{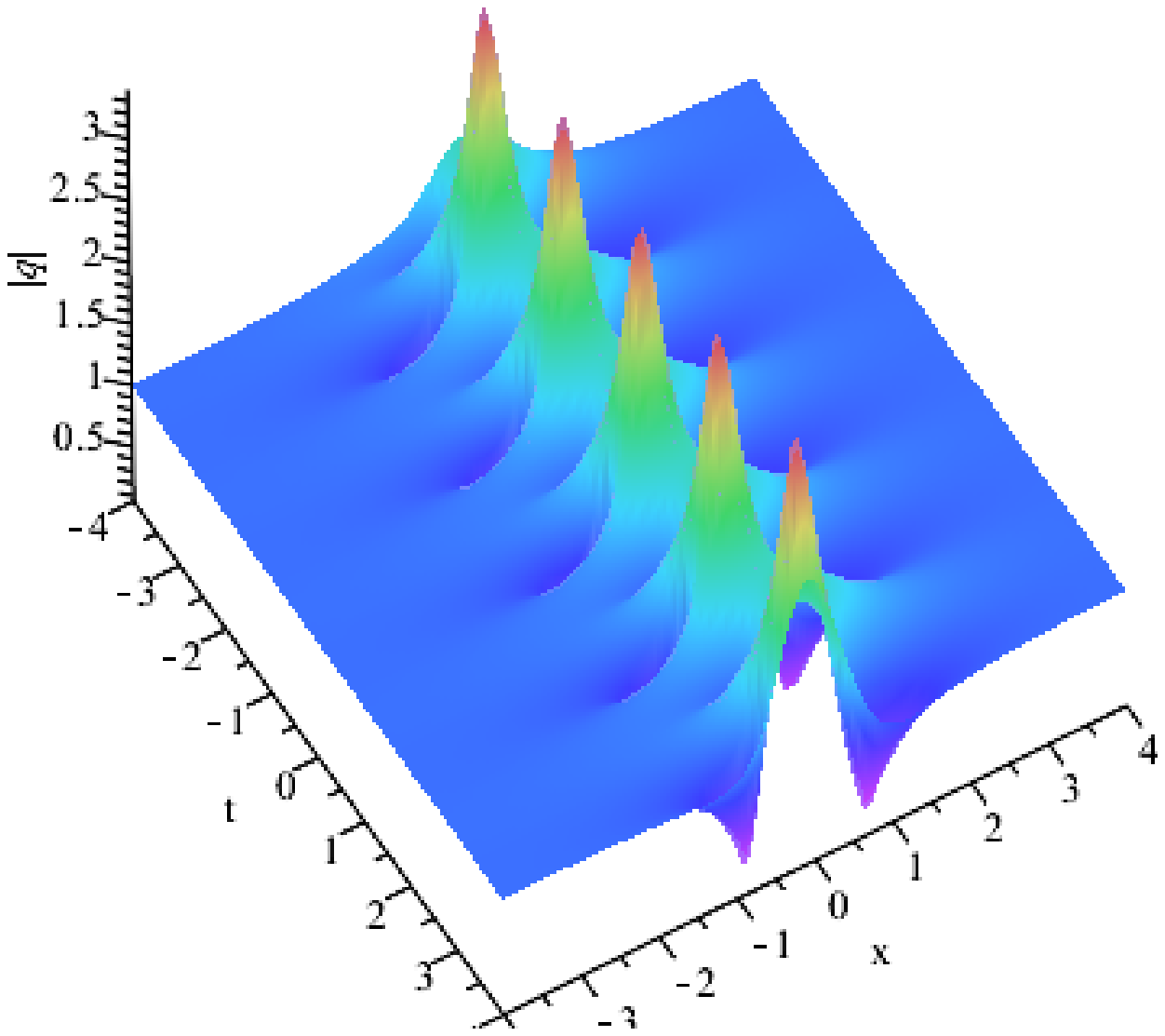}}}
\qquad\qquad\qquad\qquad
{\rotatebox{0}{\includegraphics[width=3.75cm,height=3.5cm,angle=0]{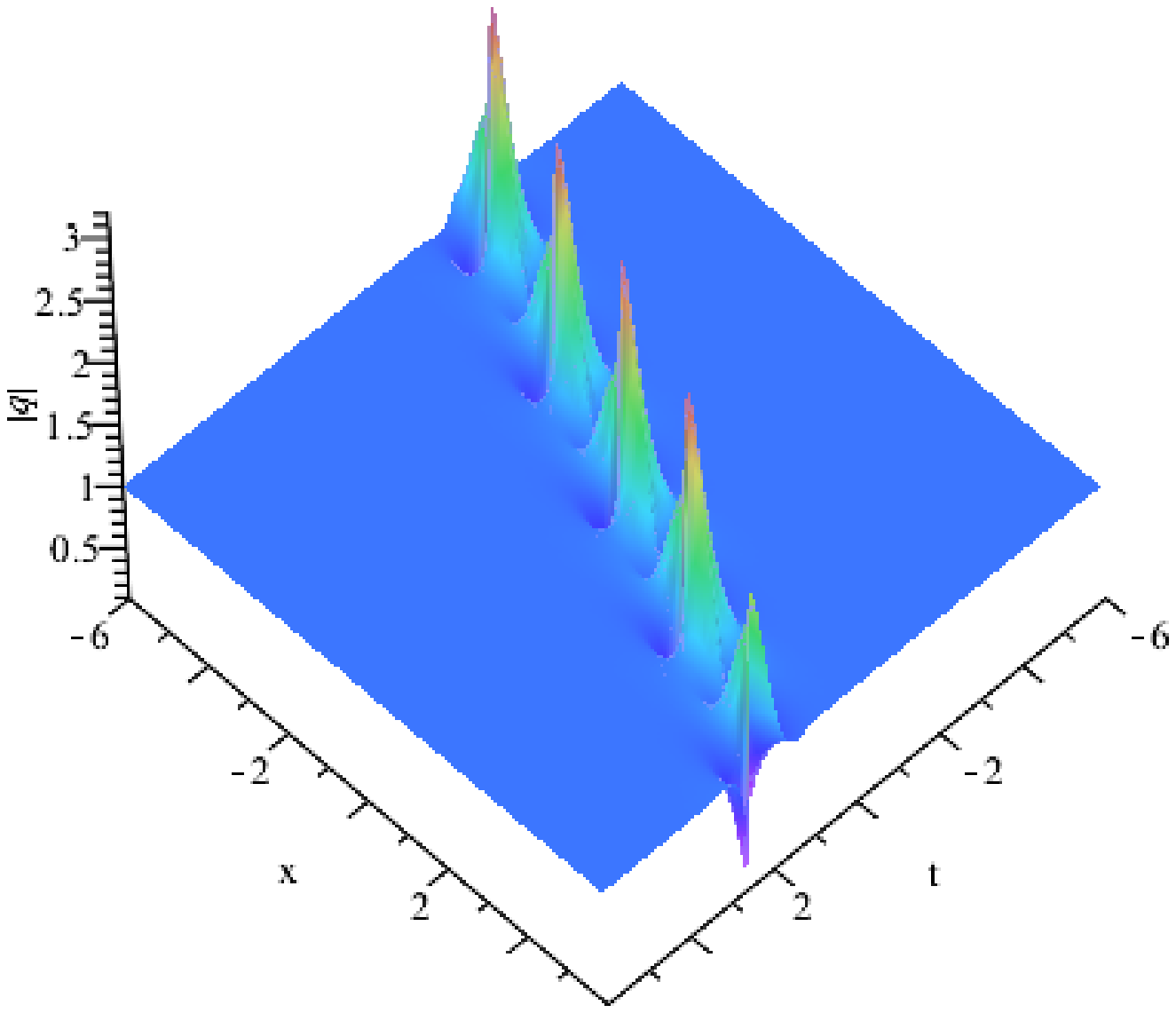}}}}

 $\qquad\qquad\qquad\quad(\textbf{a})\qquad \ \qquad\qquad\qquad\qquad\qquad \qquad\qquad\qquad(\textbf{b})$\\
\noindent {\small \textbf{Figure 3.} Propagation of the solution \eqref{Q-jie} with parameters $q_{0}=1$,  $\rho_{1,0}=1$, $\rho_{1,1}=2$, $\rho_{2,0}=2$, $\rho_{2,1}=1$, (\textbf{a}) $z_{1}=2i$, (\textbf{b}) $z_{1}=1+2i$.}
\begin{rem}
The solution is consistent with the result in \cite{Biondini-JMP2014}. It is noted that the solution is homoclinic on the $x$-axis and  the  is periodic on $t$-axis, which was found by Kuznetsov in 1977 \cite{Kuznetsov-1977}.
\end{rem}

\textbf{Case(B):} Considering the  two special cases  that the discrete eigenvalue is purely imaginary shown in Fig.$4a$, and the discrete eigenvalue is complex parameter ($z_{1}=x+iy$, $x\neq0$) shown in Fig.$4b$, we observe that when the eigenvalue is a pure imaginary,  the solution behaves as a breathe solution and exhibits periodic behavior along the $t$-axis. Now taking $N=2$, some elements can be derived from \eqref{xx1}-\eqref{xx2}
\begin{align*}
&\eta_{1,1}=iq_{-}\left(\frac{\rho_{1,1}f_{1,0}}{v_{1}}+\frac{\rho_{1,2}f_{1,2}}{v_{1}}-
\frac{\rho_{1,2}f_{1,0}}{(v_{1})^{2}}\right),~
\eta_{1,2}=\frac{iq_{-}\rho_{1,2}f_{1,0}}{v_{1}},\\
&\eta_{2,1}=iq_{-}\left(\frac{\rho_{2,1}f_{2,0}}{v_{2}}+\frac{\rho_{2,2}f_{2,2}}{v_{2}}-
\frac{\rho_{2,2}f_{2,0}}{(v_{2})^{2}}\right),~
\eta_{2,2}=\frac{iq_{-}\rho_{2,2}f_{2,0}}{v_{2}},\\
&\tilde{\eta}_{1,1}=-\rho_{1,1}^{*}f_{1,0}^{*}-\rho_{1,2}^{*}f_{1,1}^{*},~~~~~~
\tilde{\eta}_{1,2}=-\rho_{1,2}^{*}f_{1,0}^{*},\\
&\tilde{\eta}_{2,1}=-\rho_{2,1}^{*}f_{2,0}^{*}-\rho_{2,2}^{*}f_{2,1}^{*},~~~~~~
\tilde{\eta}_{2,2}=-\rho_{2,2}^{*}f_{2,0}^{*},\\
&\Theta_{1,11}=\frac{\rho_{1,1}f_{1,0}}{v_{1}-v_{1}^{*}}+\frac{\rho_{1,2}f_{1,1}}{v_{1}-v_{1}^{*}}
-\frac{\rho_{1,2}f_{1,0}}{(v_{1}-v_{1}^{*})^{2}}, ~~~~~~~~~~~\Theta_{1,21}=
\frac{\rho_{1,2}f_{1,0}}{v_{1}-v_{1}^{*}},\\
&\Theta_{1,12}=\frac{\rho_{1,1}f_{1,0}}{(v_{1}-v_{1}^{*})^{2}}+\frac{\rho_{1,2}f_{1,1}}
{(v_{1}-v_{1}^{*})^{2}}-\frac{\rho_{1,2}f_{1,0}}{(v_{1}-v_{1}^{*})^{3}}, ~~\Theta_{1,22}=
\frac{\rho_{1,2}f_{1,0}}{(v_{1}-v_{1}^{*})^{2}},\\
&\Theta_{2,11}=\frac{\rho_{1,1}f_{1,0}}{v_{1}-v_{2}^{*}}+\frac{\rho_{1,2}f_{1,1}}{v_{1}-v_{2}^{*}}
-\frac{\rho_{1,2}f_{1,0}}{(v_{1}-v_{2}^{*})^{2}}, ~~~~~~~~~~~\Theta_{1,21}=
\frac{\rho_{1,2}f_{1,0}}{v_{1}-v_{2}^{*}},\\
&\Theta_{2,12}=\frac{\rho_{1,1}f_{1,0}}{(v_{1}-v_{2}^{*})^{2}}+\frac{\rho_{1,2}f_{1,1}}
{(v_{1}-v_{2}^{*})^{2}}-\frac{\rho_{1,2}f_{1,0}}{(v_{1}-v_{1}^{*})^{3}}, ~~\Theta_{2,22}=
\frac{\rho_{1,2}f_{1,0}}{(v_{1}-v_{2}^{*})^{2}},\\
&\Theta_{3,11}=\frac{\rho_{2,1}f_{2,0}}{v_{2}-v_{1}^{*}}+\frac{\rho_{2,2}f_{2,1}}{v_{2}-v_{1}^{*}}
-\frac{\rho_{2,2}f_{2,0}}{(v_{2}-v_{1}^{*})^{2}}, ~~~~~~~~~~~\Theta_{2,21}=
\frac{\rho_{2,2}f_{1,0}}{v_{2}-v_{1}^{*}},\\
&\Theta_{3,12}=\frac{\rho_{2,1}f_{2,0}}{(v_{2}-v_{1}^{*})^{2}}+\frac{\rho_{2,2}f_{2,1}}
{(v_{2}-v_{1}^{*})^{2}}-\frac{\rho_{2,2}f_{2,0}}{(v_{2}-v_{1}^{*})^{3}}, ~~\Theta_{2,22}=
\frac{\rho_{2,2}f_{2,0}}{(v_{2}-v_{1}^{*})^{2}},\\
&\Theta_{4,11}=\frac{\rho_{2,1}f_{2,0}}{v_{2}-v_{2}^{*}}+\frac{\rho_{2,2}f_{2,1}}{v_{2}-v_{2}^{*}}
-\frac{\rho_{2,2}f_{2,0}}{(v_{2}-v_{2}^{*})^{2}}, ~~~~~~~~~~~\Theta_{4,21}=
\frac{\rho_{2,2}f_{1,0}}{v_{2}-v_{2}^{*}},\\
&\Theta_{4,12}=\frac{\rho_{2,1}f_{2,0}}{(v_{2}-v_{1}^{*})^{2}}+\frac{\rho_{2,2}f_{2,1}}
{(v_{2}-v_{2}^{*})^{2}}-\frac{\rho_{2,2}f_{2,0}}{(v_{2}-v_{2}^{*})^{3}}, ~~\Theta_{4,22}=
\frac{\rho_{2,2}f_{2,0}}{(v_{2}-v_{2}^{*})^{2}}.
\end{align*}

\centerline{{\rotatebox{0}{\includegraphics[width=3.75cm,height=3.5cm,angle=0]{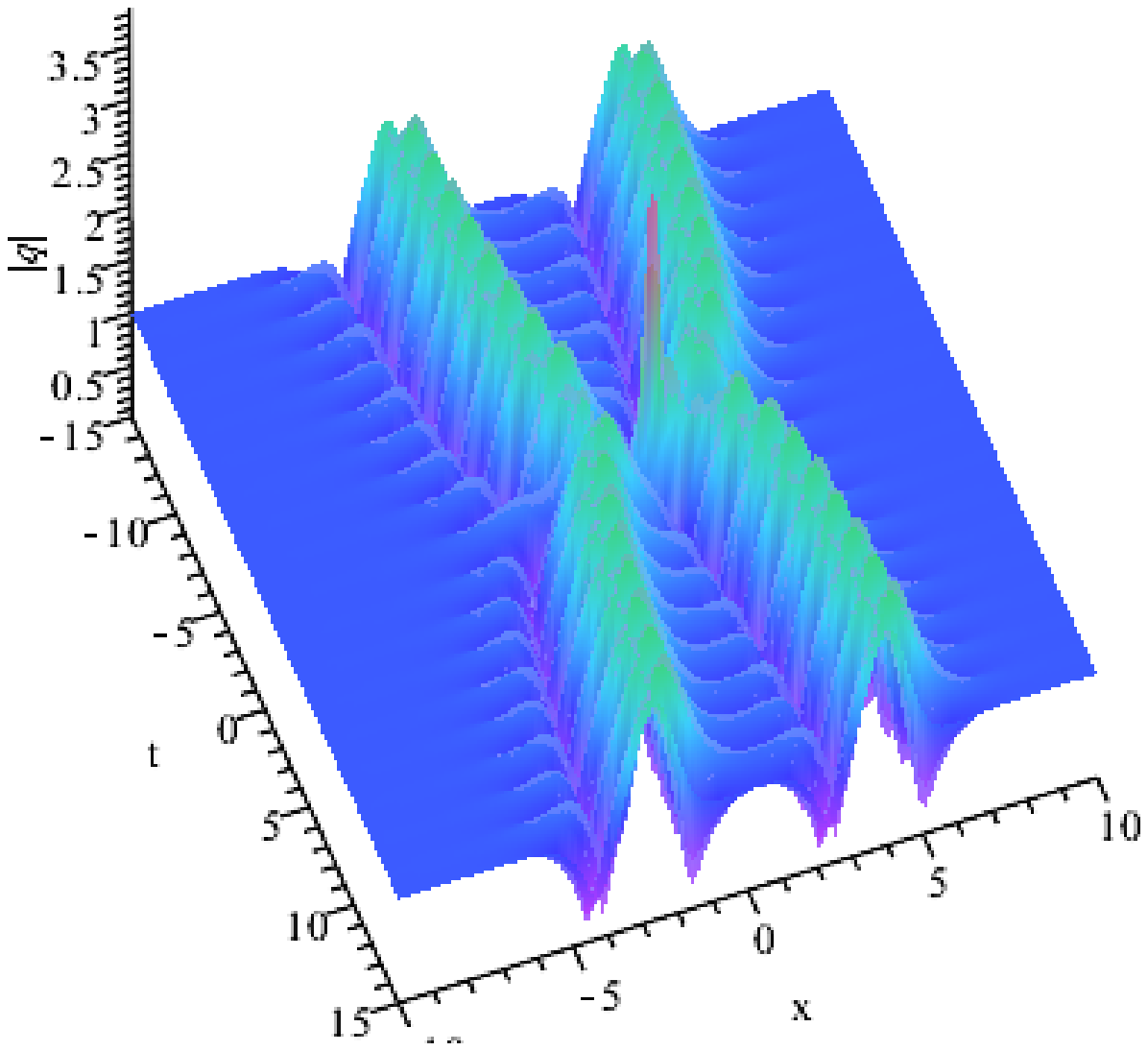}}}
\qquad\qquad\qquad\qquad
{\rotatebox{0}{\includegraphics[width=3.75cm,height=3.5cm,angle=0]{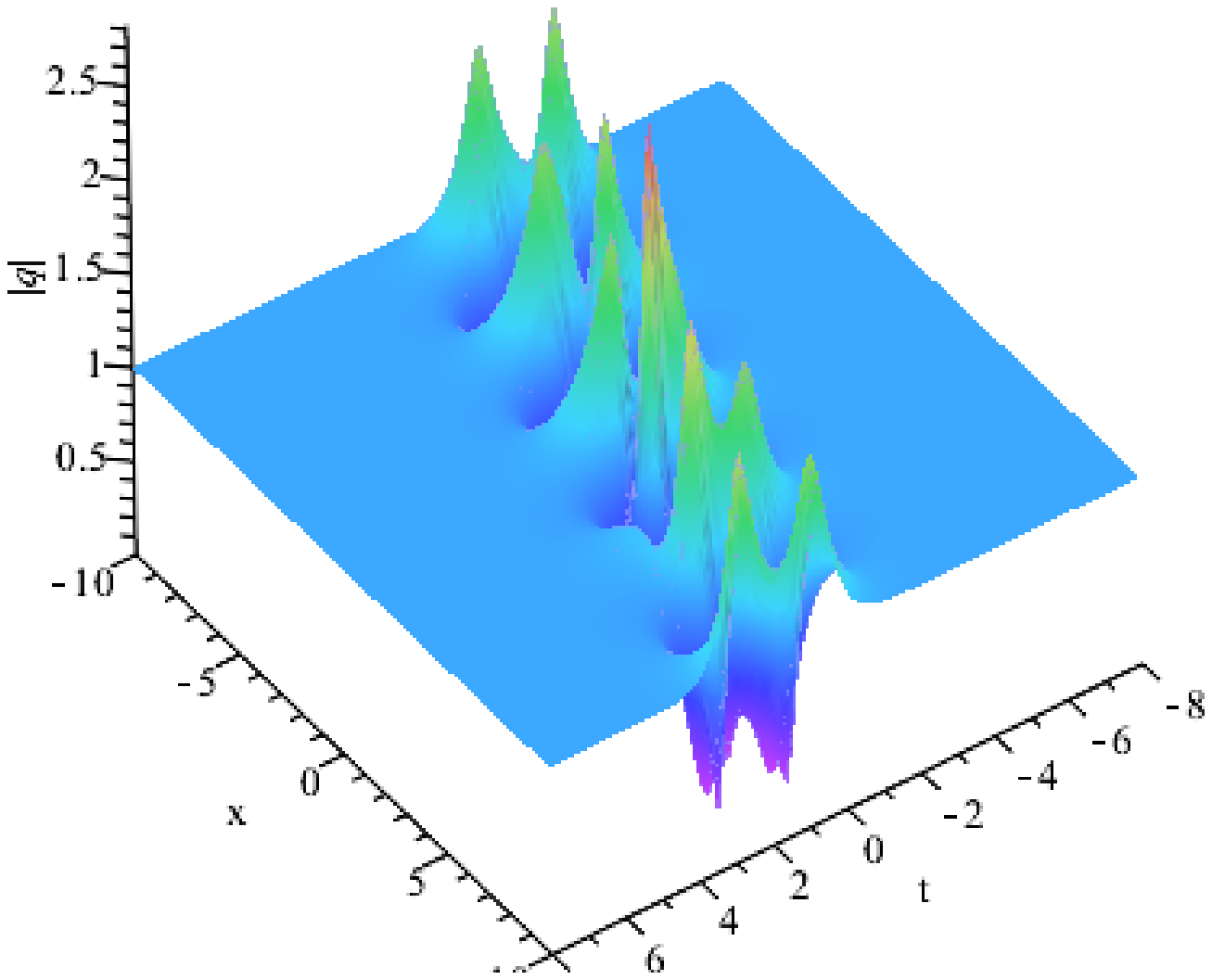}}}}

 $\qquad\qquad\qquad\quad(\textbf{a})\qquad \ \qquad\qquad\qquad\qquad\qquad \qquad\qquad\qquad(\textbf{b})$\\
\noindent {\small \textbf{Figure 4.} Propagation of the solution \eqref{Q-jie} with parameters $q_{0}=1$,  $\rho_{1,0}=1$, $\rho_{1,1}=2$, $\rho_{1,2}=1$, $\rho_{2,0}=3$, $\rho_{2,1}=1$ and $\rho_{2,2}=2$  (\textbf{a}) $z_{1}=2i$, (\textbf{b}) $z_{1}=1+i$.}
\begin{rem}
For the reflection coefficient has second-order poles, the results are consistent with those in
\cite{Pichler-2017}.
\end{rem}

\textbf{Case(C):} For a third pole, i.e., $N=3$. For the sake of brevity, we omit the elements expression of $\Theta$, $|K_{1}\rangle $ and $|K_{2}\rangle$ here and give the propagation behavior of the solution directly

\centerline{{\rotatebox{0}{\includegraphics[width=3.75cm,height=3.5cm,angle=0]{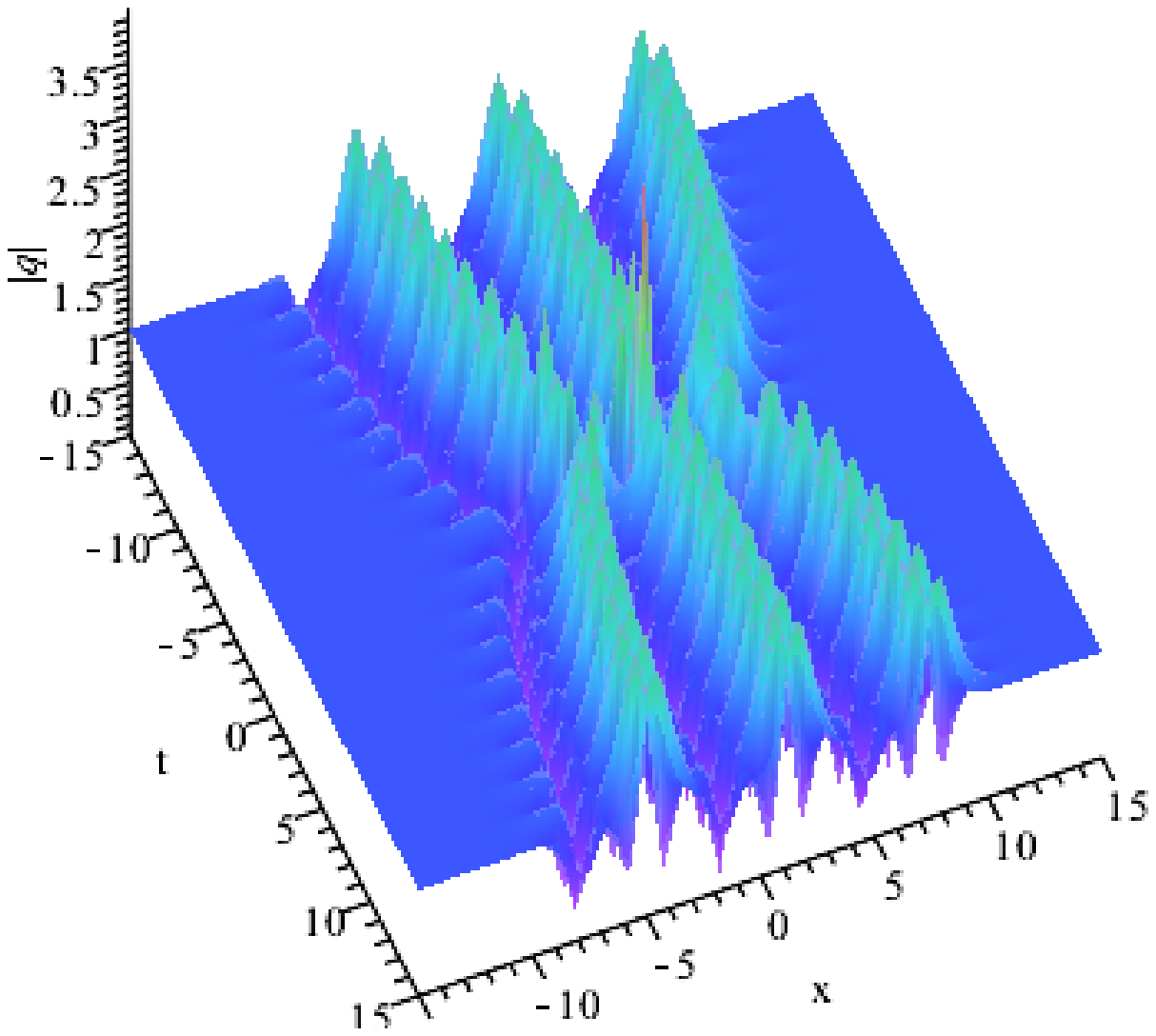}}}
\qquad\qquad\qquad\qquad
{\rotatebox{0}{\includegraphics[width=3.75cm,height=3.5cm,angle=0]{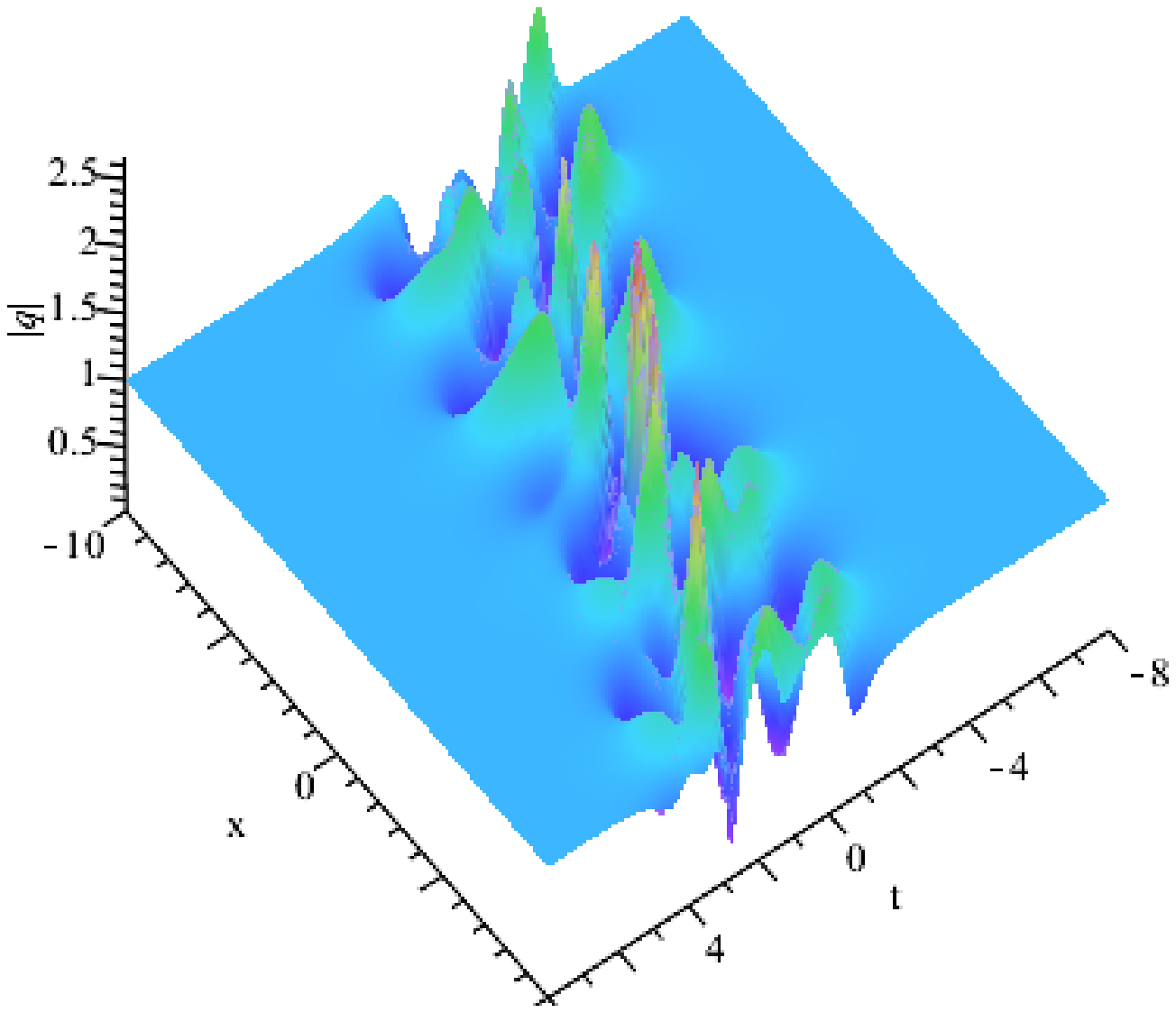}}}}

 $\qquad\qquad\qquad\quad(\textbf{a})\qquad \ \qquad\qquad\qquad\qquad\qquad \qquad\qquad\qquad(\textbf{b})$\\
\noindent {\small \textbf{Figure 5.} Propagation of the solution \eqref{Q-jie} with parameters $q_{0}=1$,  $\rho_{1,0}=1$, $\rho_{1,1}=2$, $\rho_{1,2}=2$, $\rho_{1,3}=1$, $\rho_{2,0}=2$, $\rho_{2,1}=2$, $\rho_{2,2}=1$ and $\rho_{2,3}=3$  (\textbf{a}) $z_{1}=2i$, (\textbf{b}) $z_{1}=1+i$.}

It is noted that this phenomenon has not appeared in \cite{Biondini-JMP2014,Pichler-2017}, that is, for the case that the reflection coefficient $\rho(z)$ a has the third-order poles. It is observed from the phenomenon shown in Fig. 3-5 that the corresponding results of the third-order poles when the eigenvalues are pure imaginary can be regarded as the linear superposition of the solutions corresponding to the simple poles and the second-order poles. Furthermore, we can observe the behavior of the solutions corresponding to the fourth-order poles or even higher-order poles, which is also an advantage of this work, that is, we give a unified expression of the solutions corresponding to each order pole.

\section{Multiple high-order poles}

In this section, we discuss the general case that scattering data $s_{11}(z)$ has $N$ higher-order poles $z_{1},z_{2},\ldots,z_{N}$, and the corresponding power of each pole is $n_{1},n_{2},\ldots,n_{N}$. Let $v_{1}^{k}=z_{k}$ and $v_{2}^{k}=-q_{0}^{2}/z_{k}^{*}$ ($k=1,2,\ldots,N$), then $\rho(z)$ and $\rho^{*}(z^{*})$ can be expressed as follow at each poles $z=v_{j}^{k}$ ($j=1,2$)
\begin{align}\label{4}
&\rho^{k}(z)=\rho_{0}^{k}(z)+\sum_{m_{j}=1}^{n_{k}}\frac{\rho^{k}_{1,m_{j}}}
{(z-v_{1}^{k})^{m_{j}}},\quad\quad~~
\rho^{k}(z)=\widehat{\rho}_{0}^{k}(z)+\sum_{m_{j}=1}^{n_{k}}
\frac{\rho^{k}_{2,m_{j}}}{(z-v_{2}^{k})^{m_{j}}},\\
&\rho^{k*}(z^{*})=\rho^{k*}_{0}(z^{*})+\sum_{m_{j}=1}^{n_{k}}\frac{\rho_{1,m_{j}}^{k*}}
{(z-v_{1}^{k*})^{m_{j}}},\quad
\rho^{k*}(z^{*})=\widehat{\rho}^{k*}_{0}(z^{*})+\sum_{m_{j}=1}^{n_{k}}
\frac{\rho_{2,m_{j}}^{k*}}{(z-v_{2}^{k*})^{m_{j}}},
\end{align}
where $\rho_{j,m_{j}}^{k}$ ($j=1,2$, $k=1,2,\ldots,N$ and $m_{j}=1,2,\ldots,n_{k}$) are defined by
\begin{align*}
\rho_{j,m_{j}}^{k}=\lim_{z\rightarrow v_{j}^{k}}\frac{1}{(n_{k}-n_{j})!}\frac{\partial^{n_{k}-n_{j}}}{\partial z^{n_{k}-n_{j}}}
\left[(z-v_{j}^{k})^{n_{k}}\rho_{m}(z)\right].
\end{align*}

Similar to the case of a single high-order pole, a closed algebraic system can be obtained by considering each pole separately. Then taking the following  notations
\begin{align}
&|\Xi_{1}\rangle=\left(|\Xi_{1}^{1}\rangle,|\Xi_{1}^{2}\rangle,\ldots,|\Xi_{1}^{N}\rangle\right)^{T},~~
|\Xi_{1}^{k}\rangle=\left(|\eta_{1}^{k}\rangle,|\eta_{2}^{k}\rangle\right)^{T},\\
&|\Xi_{2}\rangle=\left(|\Xi_{2}^{1}\rangle,|\Xi_{2}^{2}\rangle,\ldots,|\Xi_{2}^{N}\rangle\right)^{T},~~
|\Xi_{2}^{k}\rangle=\left(|\tilde{\eta}_{1}^{k}\rangle,|\tilde{\eta}_{2}^{k}\rangle\right)^{T},\\
&|\eta_{1}^{k}\rangle=\left(|\eta_{1,1}^{k}\rangle,|\eta_{1,2}^{k}\rangle,\ldots,
|\eta_{1,N}^{k}\rangle\right)^{T},~~~
|\eta_{2}^{k}\rangle=\left(|\eta_{2,1}^{k}\rangle,|\eta_{2,2}^{k}\rangle,\ldots,
|\eta_{2,N}^{k}\rangle\right)^{T},\\
&|\tilde{\eta}_{1}^{k}\rangle=\left(|\tilde{\eta}_{1,1}^{k}\rangle,|\tilde{\eta}_{1,2}^{k}\rangle,\ldots,
|\tilde{\eta}_{1,N}^{k}\rangle\right)^{T},~~~
|\tilde{\eta}_{2}^{k}\rangle=\left(|\tilde{\eta}_{2,1}^{k}\rangle,|\tilde{\eta}_{2,2}^{k}\rangle,\ldots,
|\tilde{\eta}_{2,N}^{k}\rangle\right)^{T},\\
&|\eta_{1,s}^{k}\rangle=iq_{-}\sum_{j=s}^{n_{k}}\sum_{\ell=0}^{j-s}\frac{(-1)^{\ell}}
{(v_{1}^{k})^{\ell+1}}r_{1,j}^{k}f_{1,j-s-\ell}^{k},~~
|\eta_{2,s}^{k}\rangle=iq_{-}\sum_{j=s}^{n_{k}}\sum_{\ell=0}^{j-s}\frac{(-1)^{\ell}}
{(v_{2}^{k})^{\ell+1}}r_{2,j}^{k}f_{2,j-s-\ell}^{k},\\
&|\tilde{\eta}_{1,s}^{k}\rangle=-\sum_{j=s}^{n_{k}}r_{1,j}^{k*}f_{1,j-s-\ell}^{k*},~~~~
|\tilde{\eta}_{2,s}^{k}\rangle=-\sum_{j=s}^{n_{k}}r_{2,j}^{k*}f_{2,j-s-\ell}^{k*},
\end{align}
in order to write the closed algebraic system into matrix form, we further introduce
\begin{align}
&\Gamma=\left(
         \begin{array}{cccc}
           (\Gamma_{11}) & (\Gamma_{12}) & \cdots & (\Gamma_{1N}) \\
           (\Gamma_{21}) & (\Gamma_{22}) & \cdots & (\Gamma_{2N}) \\
           \vdots & \vdots & \ddots & \vdots \\
           (\Gamma_{N1}) & (\Gamma_{N2}) & \cdots & (\Gamma_{NN}) \\
         \end{array}
       \right),~~(\Gamma_{j\ell})_{2n_{j}\times 2n_{\ell}}=\left(
                     \begin{array}{cc}
                       (\Gamma_{j\ell}^{(1)})_{n_{j}\times n_{\ell}} & (\Gamma_{j\ell}^{(2)})_{n_{j}\times n_{\ell}} \\
                       (\Gamma_{j\ell}^{(3)})_{n_{j}\times n_{\ell}} & (\Gamma_{j\ell}^{(4)})_{n_{j}\times n_{\ell}} \\
                     \end{array}
                   \right),\\
&\Gamma_{j\ell,pq}^{(1)}=\sum_{j=p}^{n_{j}}\sum_{s_{j}=0}^{j-p}\left(
                                      \begin{array}{c}
                                        q+s_{j}-1 \\
                                        s_{j} \\
                                      \end{array}
                                    \right)
\frac{(-1)^{s_{j}}\rho_{1,j}^{j}f_{1,j-s_{j}-p}^{j}(x,t)}{(v_{1}^{j}-v_{1}^{j*})^{s_{j}+q}},\\
&\Gamma_{j\ell,pq}^{(2)}=\sum_{j=p}^{n_{j}}\sum_{s_{j}=0}^{j-p}\left(
                                      \begin{array}{c}
                                        q+s_{j}-1 \\
                                        s_{j} \\
                                      \end{array}
                                    \right)
\frac{(-1)^{s_{j}}\rho_{1,j}^{j}f_{1,j-s_{j}-p}^{j}(x,t)}{(v_{1}^{j}-v_{2}^{j*})^{s_{j}+q}},\\
&\Gamma_{j\ell,pq}^{(3)}=\sum_{j=p}^{n_{j}}\sum_{s_{j}=0}^{j-p}\left(
                                      \begin{array}{c}
                                        q+s_{j}-1 \\
                                        s_{j} \\
                                      \end{array}
                                    \right)
\frac{(-1)^{s_{j}}\rho_{2,j}^{j}f_{2,j-s_{j}-p}^{j}(x,t)}{(v_{2}^{j}-v_{1}^{j*})^{s_{j}+q}},\\
&\Gamma_{j\ell,pq}^{(4)}=\sum_{j=p}^{n_{j}}\sum_{s_{j}=0}^{j-p}\left(
                                      \begin{array}{c}
                                        q+s_{j}-1 \\
                                        s_{j} \\
                                      \end{array}
                                    \right)
\frac{(-1)^{s_{j}}\rho_{2,j}^{j}f_{2,j-s_{j}-p}^{j}(x,t)}{(v_{2}^{j}-v_{2}^{j*})^{s_{j}+q}}.
\end{align}
\begin{thm}
Under the nonzero conditions \eqref{Q2}, the solution of focusing NLS with multiple high-order pole can be derived as the form
\begin{align}
q(x,t)=q_{-}-i\frac{\det\left(I_{\epsilon}+\Gamma^{*}\Gamma+|\Xi_{2}\rangle\langle Y_{0}|\right)-\det\left(I_{\epsilon}+\Gamma^{*}\Gamma+\Gamma^{*}|\Xi_{1}\rangle\langle Y_{0}|\right)}{\det\left(I_{\epsilon}+\Gamma^{*}\Gamma\right)},
\end{align}
where $\langle Y_{0}|=\left(\langle Y_{0}^{1}|,\langle Y_{0}^{2}|,\ldots,\langle Y_{0}^{N}|\right)^{T}$ with $\langle Y_{0}^{k}|=(1,0,\ldots,0,1,0,\ldots,0)_{1\times2n_{k}}$, and
\begin{align*}
I_{\epsilon}=\left(
                    \begin{array}{ccccc}
                      I_{n_{1}\times n_{1}} &   &   &   &   \\
                        & I_{n_{1}\times n_{1}} &  &   &   \\
                        &   & \ddots &   &   \\
                        &   &   & I_{n_{N}\times n_{N}} &   \\
                        &   &   &   & I_{n_{N}\times n_{N}} \\
                    \end{array}
                  \right).
\end{align*}
\end{thm}

\section{Conclusions}

In this work, the Riemann-Hilbert problem is developed to study the focusing nonlinear Schr\"{o}dinger equation with nonzero boundary conditions.  Two works about focusing NLS equation under NZBCs are recalled, that is, soliton solutions with simple pole and second pole of reflection coefficients are studied in \cite{Biondini-JMP2014,Pichler-2017}, respectively. However, for the case of higher-order poles, it is very complex to study the residues of each order, so it is impossible to obtain the soliton solutions corresponding to higher-order poles by using the method in \cite{Biondini-JMP2014,Pichler-2017}.

In \cite{Zhang-SAMP}, the authors investigate the regularity of the multiple higher-order
poles solitons of the NLS equation  with the zero boundary conditions. In order to solve the soliton solutions corresponding to higher-order poles under NZBCs, we make a transformation, that is, first expanding the piecewise analytic function $M(z)$ defined for the establishment of the corresponding RH problem at the reflection coefficient poles in combination with the asymptotic property, one knows that once the coefficients of the expansion of $M(z)$ at the poles are determined, the higher-order pole soliton solutions of the NLS equation under NZBCs can be obtained. Thus the problem is transformed into solving the coefficients of the expansion of $M(z)$ at the poles. Then we expand the reflection coefficient $\rho(z)$ at the corresponding pole by Laurent series and expand the exponential oscillation term $e^{\pm2i\theta(z)}$ by Taylor series. Finally, we can get a closed algebraic system to solve the coefficients of $M(z)$.

\section*{Acknowledgements}
This work was supported by the National Natural Science Foundation of China under Grant No. 11975306, the Natural Science Foundation of Jiangsu Province under Grant No. BK20181351, the Six Talent Peaks Project in Jiangsu Province under Grant No. JY-059,   and the Fundamental Research Fund for the Central Universities under the Grant Nos. 2019ZDPY07 and 2019QNA35.


\begin{thebibliography}{99}
\bibitem{GGKM}
Gardner CS, Greene JM, Kruskal MD, and Miura RM. Method for the solving
for the Korteweg-de Veries equation. \emph{Phys. Rev. Lett}. 1967;19: 1095-1097.

\bibitem{ZS-1972}
Zakharov VE, Shabat AB.  Exact theory of two-dimensional self-focusing and one-dimensional self-modulaiton of waves in nonlinear media. \emph{Sov. Phys. JETP}. 1972;34:62-69.

\bibitem{Ablowitz-PRL1973}
Ablowitz MJ, Kaup DJ, Newell AC, et al. Nonlinear-evolution equations of
physical significance.\emph{ Phys. Rev. Lett}. 1973;31: 125-127.

\bibitem{Ablowitz-SAPM1974}
Ablowitz MJ, Kaup DJ, Newell AC, et al. The inverse scattering transform Fourier analysis for nonlinear problems. \emph{Stud. Appl. Math}. 1974;53: 249-315.

\bibitem{Manakov-1974}
Manakov, SV. Nonlinear Fraunhofer diffraction, \emph{Sov. Phys. JETP}. 1974;38: 693-696.

\bibitem{Ablowitz-JMP1973}
Ablowitz MJ, Newell AC.  The decay of the continuous spectrum for solutions of
the KdV equation. \emph{J. Math. Phys}. 1973;14: 1277-1284.

\bibitem{Ablowitz-1981}
Ablowitz MJ, Segur H.  Solitons and the inverse scattering transform. SIAM, Philadelphia, 1981.

\bibitem{BC-1984}
Beals R, Coifman R.  Scattering and inverse scattering for first order systems,
\emph{Comm. Pure. Appl. Math}. 1984;37: 39-90.
\bibitem{Zhou-1989}
Zhou X.  The Riemann-Hilbert problem and inverse scattering, \emph{SIAM. J. Math.
Anal}. 1989;20:966-986.

\bibitem{Tian-jde}
Tian SF.  Initial-boundary value problems for the general coupled nonlinear Schr\"{o}dinger equation on the interval via the Fokas method. \emph{J. Differ. Equ}. 2017;262: 506-558.
\bibitem{Tian-pa2016}
Tian SF. The mixed coupled nonlinear Schr\"{o}dinger equation on the half-line via the Fokas method. \emph{Proc. R. Soc. Lond. A} 2016;472(2195): 20160588.
\bibitem{YZhang-2020}
Wang DS, Zhang DJ, Yang J: Integrable properties of the general coupled
nonlinear Schr\"{o}dinger equations. \emph{J. Math. Phys}.  2010;51: 023510.

\bibitem{Tian-jpa2017}
Tian SF. Initial-boundary value problems of the coupled modified Korteweg-de Vries equation on the half-line via the Fokas method. \emph{J. Phys. A: Math. Theor}. 2017;50(39): 395204.
\bibitem{Ma-2018}
Ma WX. Riemann-Hilbert problems and $N$-soliton solutions for a coupled mKdV system. \emph{J. Geom.  Phys.} 2018;132: 45-54.
\bibitem{NLiu-2019}
Liu N, Guo BL, Wang DS, and Wang YF. Long-time asymptotic behavior for an extended modified Korteweg-de Vries equation. \emph{Commun. Math. Sci}. 2019;17: 1877-1913.

\bibitem{tian-PA2018}
Tian SF, Zhang TT. Long-time asymptotic behavior for the Gerdjikov-Ivanov type of derivative nonlinear Schr\"{o}dinger equation with time-periodic boundary condition. \emph{Proc. Amer. Math. Soc}. 2018;146: 1713-1729.

\bibitem{Geng-2016}
Geng X, Wu J. Riemann-Hilbert approach and $N$-soliton solutions for a generalized Sasa-Satsuma equation. \emph{Wave. Motion}.  2016;60: 62-72.
\bibitem{Xu-2013}
Xu J, Fan E. The unified transform method for the Sasa-Satsuma equation on the half-line. \emph{Proc. R. Soc. A} 2013;469: 20130068.

\bibitem{Ma-1979}
Ma YC. The perturbed plane-wave solutions of the cubic Schr\"{o}dinger equation. \emph{Stud. Appl.  Math}. 1979;60: 43-58.


\bibitem{Biondini-JMP2014}
Biondini G, Kova\u{c}i\u{c} G. Inverse scattering transform for the focusing nonlinear Schr\"{o}dinger equation with nonzero boundary conditions. \emph{J. Math. Phys}.  2014;55(3): 031506.
\bibitem{Zhang-2020}
Zhang GQ, Yan ZY. Focusing and defocusing mKdV equations with nonzero boundary
conditions: Inverse scattering transforms and soliton interactions. \emph{Phys. D} 2020;410: 132521.
\bibitem{Zhang-2020-non}
Zhang GQ, Yan ZY. Inverse scattering transforms and soliton solutions of focusing and
defocusing nonlocal mKdV equations with non-zero boundary conditions. \emph{Phys. D} 2020;402(15): 132170.
\bibitem{Yang-2021}
Yang YL, Fan EG. Riemann-Hilbert approach to the modified nonlinear Schr\"{o}dinger equation with non-vanishing asymptotic boundary conditions. \emph{Phys. D} 2021;417: 132811.
\bibitem{Wll}
Wen LL, Fan EG. The Sasa-Satsuma equation with non-vanishing boundary conditions. arXiv:1911.11944.
\bibitem{Yang-TMP}
Yang JJ, Tian SF. Riemann-Hilbert problem for the modified Landau-Lifshitz equation with nonzero boundary conditions. \emph{Theor. Math. Phys}.  2020;205: 1611-1637.
\bibitem{GI-2020}
Zhang ZC, Fan EG. Inverse scattering transform for the Gerdjikov-Ivanov equation with nonzero boundary conditions, \emph{Z. Angew. Math. Phys.}  2020;71:149.
\bibitem{GI-NZBC}
Zhang ZC, Fan EG. Inverse scattering transform and multiple high-order pole solutions for the Gerdjikov-Ivanov equation under the zero/nonzero background, arXiv:2012.13654.
\bibitem{Pichler-2017}
Pichler M, Biondini G. On the focusing non-linear Schr\"{o}dinger equation with non-zero boundary conditions and double poles. \emph{IMA J. Appl. Math}. 2017;82(1): 131-151.

\bibitem{Wadati-1982}
Wadati M, Ohkuma K. Multiple-pole solutions of the modified Korteweg-de Vries equation. \emph{J. Phys. Soc. Jpn}. 1982;51:2029-2035.

\bibitem{Tsuru-1984}
Tsuru H, Wadati M. The multiple pole solutions of the sine-Gordon equation. \emph{J. Phys. Soc. Jpn}. 1984;53:2908-2921.

\bibitem{Gagnon-1994}
Gagnon L, Sti\'{e}venart N. $N$-soliton interaction in optical fibers: the multiple-pole case. \emph{Opt. Lett}. 1994;19:619-621.
\bibitem{Shchesnovich-JMP}
Shchesnovich VS, Yang J. General soliton matrices in the Riemann-Hilbert problem for integrable nonlinear equations. \emph{J. Math. Phys}. 2003;44:4604-4639.
\bibitem{Shchesnovich-SAMP}
Shchesnovich VS, Yang J. Higher-order solitons in the $N$-wave system. \emph{Stud. Appl. Math}. 2003;110:297-332.

\bibitem{Zhang-SAMP}
Zhang YS, Tao XX, Yao TT, He JS. The regularity of the multiple higher-order
poles solitons of the NLS equation. \emph{Stud. Appl. Math}. 2020;1-16.

\bibitem{Zhang-2019}
Zhang YS, Rao JG, Cheng Y, He JS. Riemann-Hilbert method for the Wadati-Konno-Ichikawa equation:
$N$ simple poles and one higher-order pole. \emph{Phys. D} 2019;399: 173-185.

\bibitem{Kuznetsov-1977}
Kuznetsov EA. Solitons in a parametrically unstable plasma. \emph{Sov. Phys. Dokl}. 1977;22: 507-508.





\end{thebibliography}
\end{document}